\documentclass{article}
\usepackage{arxiv}
\usepackage{graphicx} 
\usepackage{microtype}
\usepackage{subfigure}
\usepackage{booktabs} 
\usepackage{hyperref}

\usepackage{amsmath}
\usepackage{amssymb}
\usepackage{mathtools}
\usepackage{amsthm}
\usepackage{todonotes}
\usepackage{wrapfig}

\usepackage[capitalize,noabbrev]{cleveref}

\theoremstyle{plain}

\theoremstyle{definition}

\theoremstyle{remark}

\usepackage[numbers]{natbib}

\usepackage[T1]{fontenc}
\usepackage{booktabs}
\usepackage{multirow}
\usepackage{amssymb}
\usepackage{pifont}
\usepackage{lipsum}
\usepackage{tikz}
\usepackage{subfigure}

\usepackage{algorithm}
\usepackage{CJKutf8}

\usepackage{authblk}

\title{EPD-Serve: A Flexible Multimodal EPD Disaggregation Inference Serving System On Ascend}

\author[1]{Fan Bai}
\author[1]{Pai Peng}
\author[1]{Zhengzhi Tang}
\author[1]{Zhe Wang}
\author[1]{Gong Chen}
\author[1]{Xiang Lu}
\author[1]{Yinuo Li}
\author[1]{Huan Lin}
\author[1]{Weizhe Lin}
\author[1]{Yaoyuan Wang}
\author[1*]{Xiaosong Li}

\affil[1]{Huawei Technologies Co., Ltd}

\begin{document}
\begin{CJK}{UTF8}{gbsn}

\maketitle
\thispagestyle{fancy}
\footnotetext{\textsuperscript{*} Corresponding author: lixiaosong20@huawei.com}

\begin{abstract}
With the widespread deployment and continuous evolution of large models, the ability to process multiple modalities including text, image, audio, and video, has become increasingly essential for real-world applications. However, existing multimodal inference systems generally adopt a monolithic architecture, where the Encode, Prefill, and Decode stages are tightly coupled and executed on homogeneous hardware resources. This design fails to account for the heterogeneous computational characteristics of different inference stages, leading to suboptimal hardware utilization and severely constrained system throughput.
To address these limitations, we propose \emph{EPD-Serve}, a stage-level disaggregated inference serving system for multimodal models. Leveraging the computational characteristics of the inference pipeline, \emph{EPD-Serve} decomposes the end-to-end inference process into three independent stages: Encode, Prefill, and Decode.
This stage-level decoupling enables logical isolation and flexible co-located deployment of inference tasks through dynamic orchestration.
Specifically, leveraging the Ascend hardware interconnect topology, \emph{EPD-Serve} implements an asynchronous feature prefetching mechanism for the E-P stage and a hierarchical grouped KV cache transmission mechanism for the P-D stage, thereby improving cross-node communication efficiency. 
Moreover, to accommodate the properties of multimodal inputs, \emph{EPD-Serve} incorporates multi-route scheduling strategies and instance-level load balancing schemes, along with multi-stage hardware resource co-location and spatial multiplexing at the physical layer.
To validate the efficacy of \emph{EPD-Serve}, 
we conduct comprehensive experiments on a set of multimodal understanding models, 
evaluating the system's end-to-end throughput under different disaggregated deployments and SLO constraints. 
Experimental results demonstrate that, in high-concurrency multimodal scenarios,
\emph{EPD-Serve} improves throughput by 57.37-69.48\% relative to PD-disaggregated deployment while satisfying strict SLO constraints, including TTFT below 2000 ms and TPOT below 50 ms. These findings suggest a promising new direction for optimizing the architecture of multimodal large language model inference systems.
\end{abstract}

\section{Introduction}
\label{sec:intro}
\begin{figure}[htbp]
    \begin{center}
\centerline{\includegraphics[width=0.8\columnwidth]{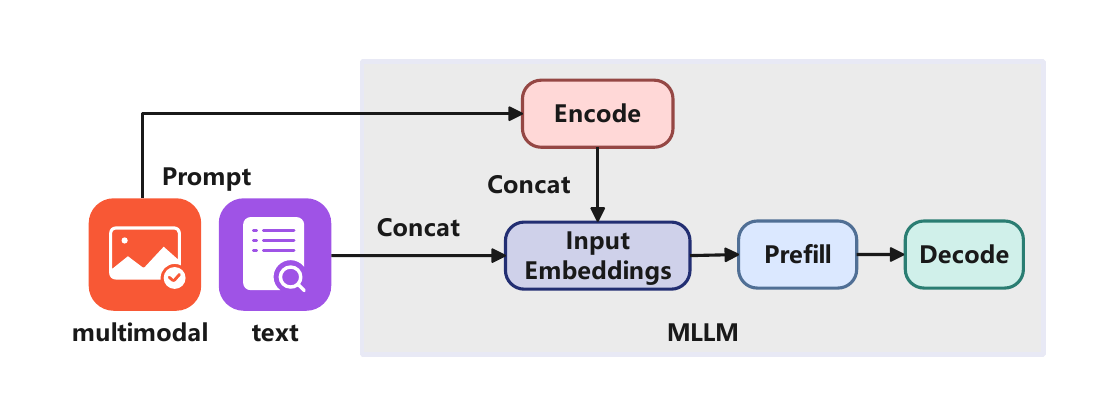}}
    \caption{Inference flow for large multi-modal language models.}
    \label{img:1_1}
    \end{center}
\end{figure}

\begin{figure}[htbp]
    \begin{center}
\centerline{\includegraphics[width=0.8\columnwidth]{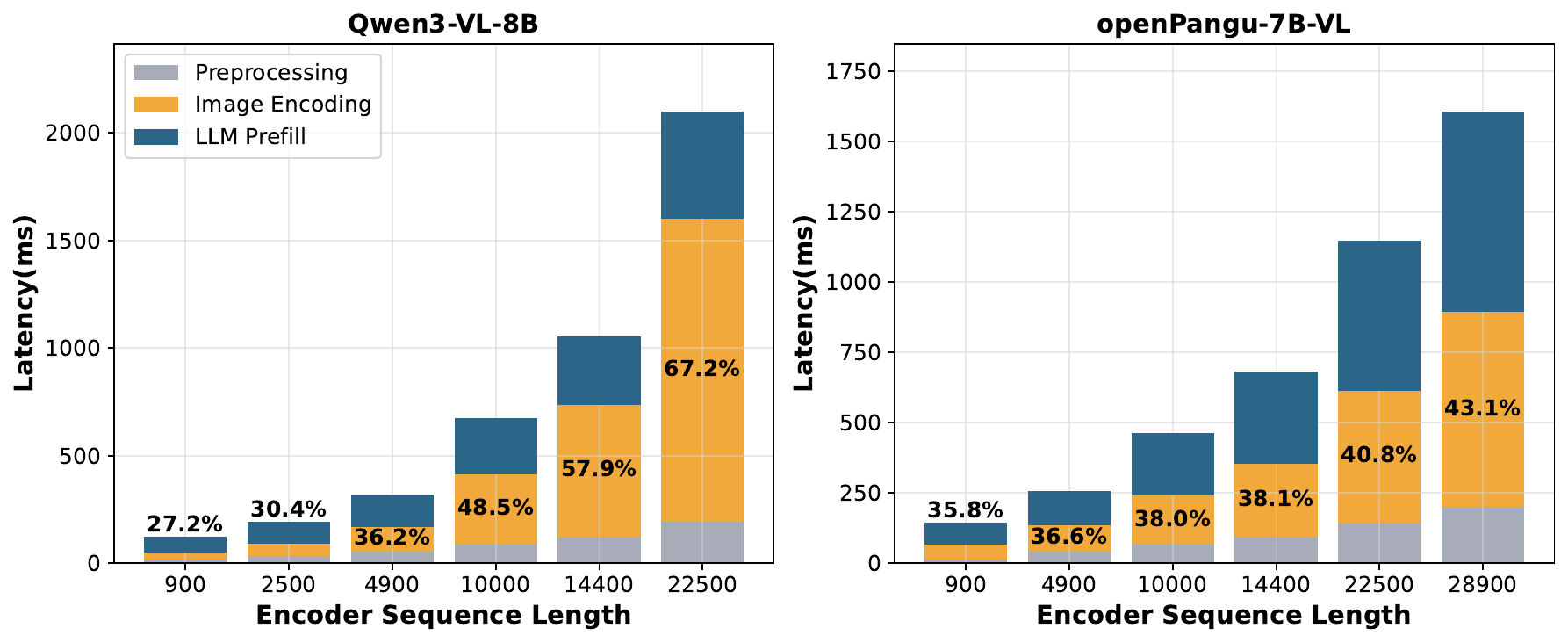}}
    \caption{Latency proportion of mainstream MLLMs as encoder sequence length increases.}
    \label{img:1_2}
    \end{center}
\end{figure}

Multimodal Large Language Models (MLLMs) \cite{radford2021learning,wu2023visual,yang2023mm} achieve cross-modal semantic alignment based on a unified language foundation, and possess the capability of comprehensive inference on multimodal inputs such as images, audio, and video. Most mainstream MLLMs adopt an architectural paradigm that couples modality encoders with a large autoregressive decoder.
Figure~\ref{img:1_1} illustrates the end-to-end inference pipeline of such models.
In typical MLLM architectures, Vision encoders, typically implemented with Vision Transformers (ViTs) containing hundreds of millions to billions of parameters, generate visual token sequences that are significantly longer than those processed by foundation language models, whose parameter scales range from billions to tens of billions, as shown in Table \ref{tab:1_1}. 
Since attention complexity grows quadratically with sequence length, the visual encoding stage can dominate end-to-end inference latency,
in some cases exceeding the Prefill time of LLM, as shown in Figure \ref{img:1_2}.
Moreover, multimodal requests exhibit wide variability in input modalities and token lengths, causing the performance bottleneck to shift dynamically across inference stages.

\begin{table}[t]
    \centering
    \footnotesize
    \setlength{\tabcolsep}{4.5pt}
    \caption{Parameter sizes of mainstream multimodal models.}
    \small
    \begin{tabular}{cccc}
    \toprule
    Models              & openPangu-7B-VL & Qwen3-VL-8B \cite{bai2025qwen3vltechnicalreport} & InternVL3-78B \cite{zhu2025internvl3exploringadvancedtraining} \\
    \midrule
    The Params   of ViT & 0.7B            & 0.6B        & 6B            \\
    The Params   of LLM & 7B              & 8B          & 72B           \\
    \bottomrule
\end{tabular}
    \label{tab:1_1}
\end{table}

Deploying such heterogeneous inference pipelines in real-world systems is challenging. Multimodal inference consists of three logically distinct stages, including Encode, Prefill, and Decode, whose data characteristics, module types, and parallelism requirements differ substantially. These differences manifest as three forms of heterogeneity. \textbf{Data heterogeneity} stems from diverse modality formats and dynamically varying sequence lengths, complicating unified dataflow management and resource allocation. \textbf{Model heterogeneity} arises because visual encoders rely on ViT or CNN modules with encoding, while LLM decoders depend on autoregressive generation with different computation and memory  characteristics, complicating system-level scheduling. \textbf{Computational heterogeneity} further separates text-only from multimodal requests, as only the latter require executing the full Encode-Prefill-Decode pipeline, causing imbalanced workloads and execution blocking.

Existing inference frameworks, such as vLLM \cite{kwon2023efficient}, SGLang \cite{zheng2024sglang}, TGI \cite{aminabadi2022deepspeed}, extend language-centric designs to multimodal settings by tightly coupling Encode and Prefill on the same hardware resources. 
This monolithic architecture exposes three critical performance bottlenecks in high-concurrency scenarios. First, \textbf{stage coupling creates execution interference}: visual encoding and text prefill compete for shared resources without isolation, allowing multimodal requests to block text-only requests, inflating Time-To-First-Token (TTFT), and disrupting Decode scheduling, thereby degrading Time-Per-Output-Token (TPOT) and overall throughput. Second, \textbf{a unified parallelism strategy fails to accommodate heterogeneous stage requirements}: Encode prefers data or sequence parallelism, whereas Decode benefits from tensor parallelism for latency reduction. Unified parallelism prevents stage-specific optimization and limits scalability. Third, \textbf{strict serial execution prevents resource reuse}: Encode, Prefill, and Decode run exclusively in sequence despite complementary compute-memory characteristics, leaving substantial NPU resources underutilized.

These limitations render monolithic multimodal inference increasingly inefficient as deployment scales and hybrid-modal traffic intensifies.
To overcome this fundamental architectural mismatch, we propose \emph{EPD-Serve}, a flexible inference serving system that decouples the pipeline into independently schedulable Encode, Prefill, and Decode stages. 
\emph{EPD-Serve} supports flexible disaggregation and co-location strategies, including E-P-D, EP-D, ED-P, and E-PD, enabling time-division specialization and spatial multiplexing tailored to diverse workload patterns. 
To mitigate the communication overhead introduced by disaggregation, we further design (1) an asynchronous feature prefetching mechanism that overlaps E-P data transfer with computation, and (2) a hierarchical grouped KV cache transmission mechanism that reduces and defers Prefill–Decode KV transfers.
We evaluate \emph{EPD-Serve} using the openPangu-7B-VL model and the ShareGPT-4o workload. Under an average load of 12 requests per second per NPU, \emph{EPD-Serve} improves throughput by 57.37-69.48\% over a strong PD-disaggregated baseline while satisfying stringent Service Level Objectives (SLOs) with $TTFT \leq 2000 ms$ and $TPOT \leq 50 ms$. These results demonstrate that EPD disaggregation is a principled and effective architectural direction for high-performance multimodal inference.

\section{Related Work}
\subsection{Prefill-Decode Disaggregation}
Prefill-Decode (PD) disaggregation, which deploys the Prefill and Decode stages on separate compute units, has become an effective approach for improving the inference performance of large language models (LLMs). The Prefill and Decode stages exhibit significant discrepancies in computational intensity and memory requirements. When deployed monolithically, these differences lead to resource contention and scheduling interference, degrading throughput and latency. Disaggregated deployment allows differentiated resource allocation and stage-specific optimization, thereby enabling parallel execution, reducing interference, and improving the throughput for the overall system.

Recent systems such as Splitwise \cite{patel2024splitwise}, DistServe \cite{zhong2024distserve}, MemServe \cite{hu2024memserve}, 
and Mooncake \cite{qin2025mooncake} propose PD-disaggregated inference architectures for LLMs. 
By introducing fine-grained resource scheduling and stage-level optimization mechanisms, 
these systems effectively mitigate cross-stage interference and significantly reduce TPOT, 
improving performance and stability. However, multimodal large model inference requires an additional compute-intensive Encode stage to convert raw inputs such as images, video, and audio into unified sequence features. This breaks the traditional two-stage inference paradigm, revealing limitations in existing PD disaggregation designs while creating new opportunities for advancing disaggregated inference systems.

\subsection{Encode-Prefill-Decode Disaggregation}
With increasingly diverse user interaction modalities, including image, video, audio and actions, the demand for multimodal large model inference has grown rapidly. Multimodal models typically introduce an additional Encoder module to transform raw perceptual inputs into token sequences aligned with the input space of the language model. While beneficial for multimodal understanding, this stage adds significant computational overhead and new data dependencies, which complicate resource allocation and scheduling in traditional inference architectures.

To address these challenges, recent work extends disaggregation beyond the PD stage to a fully disaggregated Encode-Prefill-Decode (EPD) architecture. 
Systems including HydraInfer \cite{dong2025hydrainfer} and SpaceServe \cite{zhangspaceserve} deploy the three stages independently to eliminate resource interference from the Encode stage and improve concurrency and scalability. 
ModServe \cite{qiu2025modserve} separates image and text inference pipelines and applies modality-aware routing to improve heterogeneous resource utilization. 
Additionally, HydraInfer supports dynamically matching the coupling modes and number of nodes for stages E, P, and D according to task workloads and resource status, achieving an adaptive balance. However, current EPD systems focus mainly on logical stage disaggregation and scheduling strategies. 
Their compute resources are still deployed in an exclusive-device manner, and the impact of multi-stage physical co-location on system performance has not been systematically explored. 
In contrast, this work investigates physical co-location and spatial multiplexing under the EPD disaggregation architecture, enabling flexible resource sharing across Encode, Prefill, and Decode and improving throughput, latency, and SLO attainment rate.

\section{Method}
\label{method}
This chapter presents the design of \emph{EPD-Serve}, a multimodal EPD-disaggregated inference system. 
First, an overview of the overall EPD-disaggregated framework is provided in Section \ref{epd_serve_system}. 
Next, we focus on the tensor transmission optimization mechanisms of the disaggregated system, including the E-P disaggregated transmission in Section \ref{ep_Transmission} and the P-D disaggregated transmission in Section \ref{pd_Transmission}. 
Subsequently, the system's scheduling strategy is introduced in Section \ref{modality_multi_path}. 
Finally, Section \ref{logical_physical} discusses the throughput improvements and resource efficiency enabled by flexible co-location and stage disaggregation.

\subsection{EPD-Serve System}
\label{epd_serve_system}
\begin{figure}[htbp]
    \begin{center}
\centerline{\includegraphics[width=1.0\columnwidth]{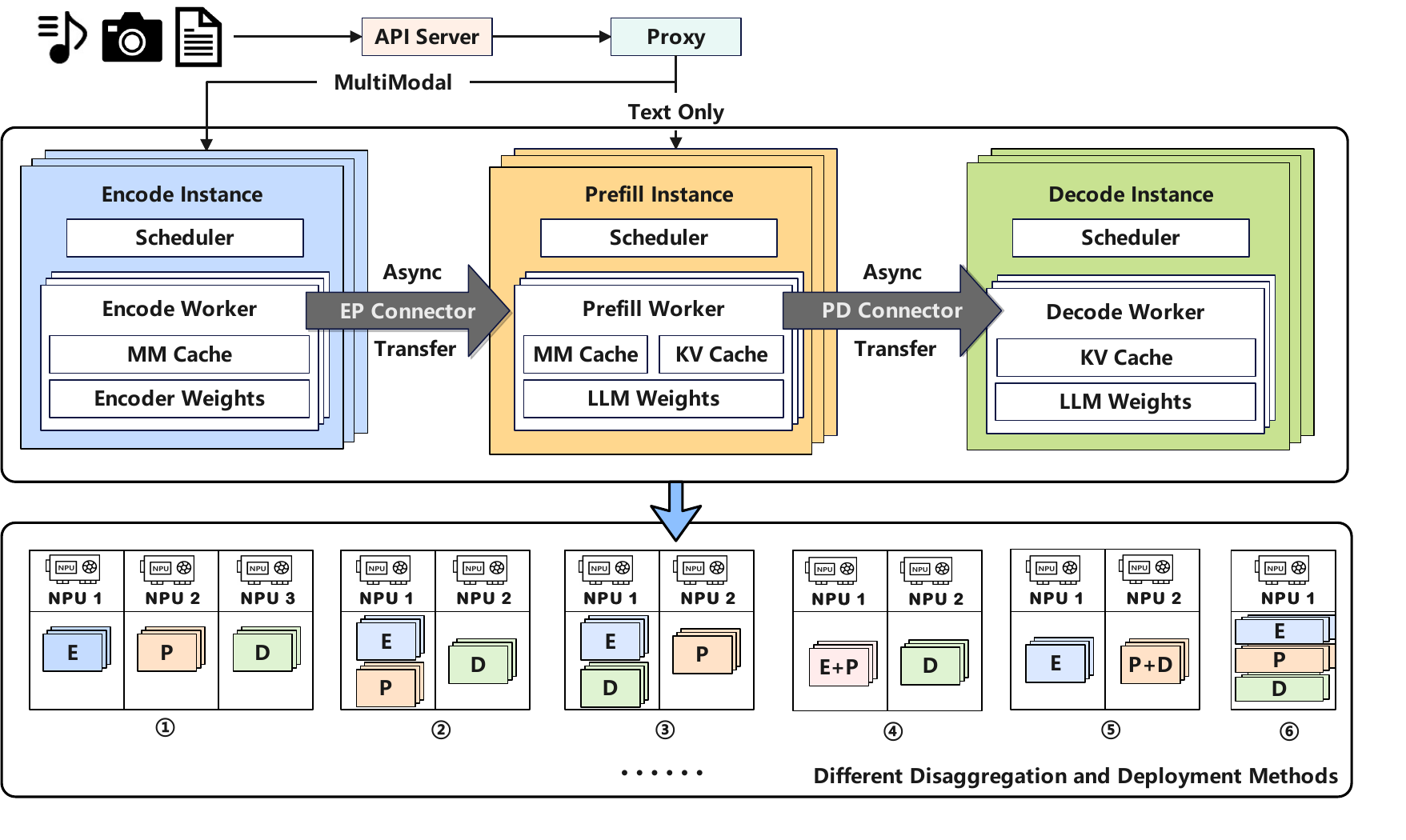}}
    \caption{Overview of the \emph{EPD-Serve} disaggregated inference architecture.}
    \label{img:3_1}
    \end{center}
\end{figure}

MLLMs typically execute inference through three stages: Encode, Prefill, and Decode. 
In traditional monolithic architectures, these stages run sequentially on the same computing resources, which fails to accommodate their significant computational heterogeneity. 
To overcome this limitation, we propose \emph{EPD-Serve}, an inference architecture based on stage-level disaggregation. 
By modularizing the inference pipeline, \emph{EPD-Serve} enables independent scheduling, efficient cross-stage transmission, and flexible resource co-location. 
The overall system architecture is shown in Figure\ref{img:3_1}.

In \emph{EPD-Serve}, the inference pipeline of MLLMs is partitioned into three stages: Encode, Prefill, and Decode. The functionality and data flow of each stage are outlined below.

\paragraph{Encode Stage:}
This stage processes multimodal inputs $I_m$ such as images, audio and video through the multimodal encoder $E$, producing high-dimensional embedding sequences:
\begin{equation}
V_m = E(I_m)
\end{equation}
where $ V_m \in \mathbb{R}^{n \times d} $ represents $n$ multimodal tokens, each with a $d$-dimensional feature vector. 
These embeddings serve as input features for the Prefill stage.

\paragraph{Prefill Stage:}
Text prompts $I_t$ are encoded into a text feature sequence $V_t$. 
The multimodal and text features are concatenated and fed into the Prefill stage:
\begin{equation}
O_1, \text{KV}_1 = P(V_m, V_t)
\end{equation}
which generates the first output token $O_1$ and constructs the KVCache $\text{KV}_1$, providing contextual dependencies for the Decode stage.

\paragraph{Decode Stage:}
Decode autoregressively produces subsequent tokens. Given the previous token and KVCache:
\begin{equation}
O_2, \text{KV}_2 = P(O_1, \text{KV}_1)
\end{equation}
this process iterates until the end-of-sequence token \texttt{<eos>} appears or a maximum generation length is reached. Due to its strict dependency on historical context, Decode operates sequentially.

\paragraph{Cross-Stage Transmission:}
\emph{EPD-Serve} enables efficient tensor transmission through asynchronous cross-stage transmission modules:
E-P transmission supports cross-node migration of multimodal feature vectors with asynchronous transmission and cache prefetching to reduce stage switching latency. 
P-D transmission uses hierarchical grouped caching and delayed transmission to precisely overlap communication with Prefill computation, enabling efficient transfer of KVCache from Prefill to Decode stages.

Overall, the EPD-disaggregated architecture achieves pipelined and parallelized inference through stage-level disaggregation and asynchronous communication. This design eliminates cross-stage resource interference, improves spatiotemporal utilization of compute and memory resources, and significantly reduces latency jitter and throughput degradation under high concurrency, thereby improving elastic scalability and maintaining stable SLO performance.

\subsection{Transmission Mechanism for E-P Disaggregation}
\label{ep_Transmission}
\begin{figure}[htbp]
    \begin{center}
\centerline{\includegraphics[width=1.0\columnwidth]{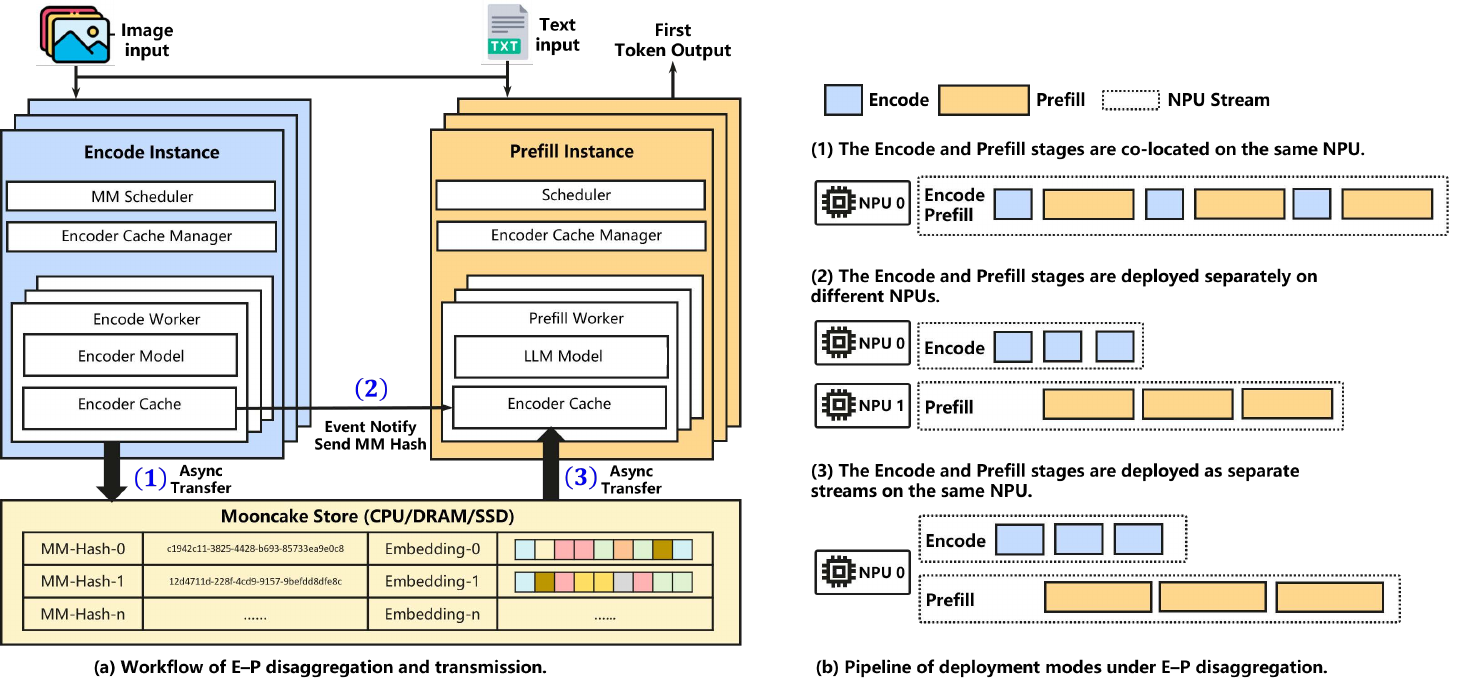}}
    \caption{E-P disaggregation with asynchronous feature transmission and pipelining. 
    Left: Asynchronous feature transmission between the Encode and Prefill stages via the Mooncake Store \cite{qin2025mooncake}. 
    Right: Pipeline layouts supporting standalone deployment or physical co-location with the Prefill stage.}
    \label{img:3_2}
    \end{center}
\end{figure}

As shown in Figure \ref{img:3_2}, in the E-P disaggregated architecture, multimodal features generated by the Encode stage are transmitted to Prefill instances, and the associated transmission latency directly affects TTFT, becoming a primary bottleneck. To mitigate this issue, we propose an event-driven asynchronous feature prefetching mechanism that enables low-blocking transmission via three optimizations: cache pooling, asynchronous prefetching, and fault-tolerant recomputation.

\paragraph{Multimodal Cache Pooling:}
\emph{EPD-Serve} implements a shared multimodal cache pool, named MM Store, that stores encoded multimodal features using the hash of multimodal inputs as the key and the corresponding feature vectors as the value. This design avoids duplicate caching and transmission, supports cross-request reuse of features, and improves cache utilization and hit rate.

\paragraph{Event-Driven Asynchronous Prefetching:}
To overlap transmission latency, feature transfer is decoupled from computation using an event-driven mechanism. After the Encode stage completes, only the feature hash is asynchronously sent to the target Prefill instance. Upon receiving the event, the listener in the Prefill stage retrieves the feature from the MM Store using the hash and writes it to the local cache. By transmitting only lightweight hash values and parallelizing execution and data transfer, this approach significantly reduces TTFT.

\paragraph{Fault-Tolerant and Recomputation:}
To enhance robustness, if a Prefill instance fails to retrieve the feature from the MM Store, the system triggers local recomputation to regenerate the missing vector. This mechanism preserves pipeline continuity while balancing performance and reliability.

\subsection{Transmission Mechanism for P-D Disaggregation}
\label{pd_Transmission}
In the P-D disaggregated architecture, the large KVCache generated during Prefill must be transmitted to Decode instances. Under synchronous transmission, all KV data across layers must be transferred once Prefill computation completes, which leads to communication congestion and significantly increases TTFT. To mitigate this overhead, we propose a hierarchically grouped KV transmission mechanism that aligns KV transfer with the model's computation pipeline, enabling parallel transmission and Prefill computation.

\begin{figure}[htbp]
    \begin{center}
\centerline{\includegraphics[width=1.0\columnwidth]{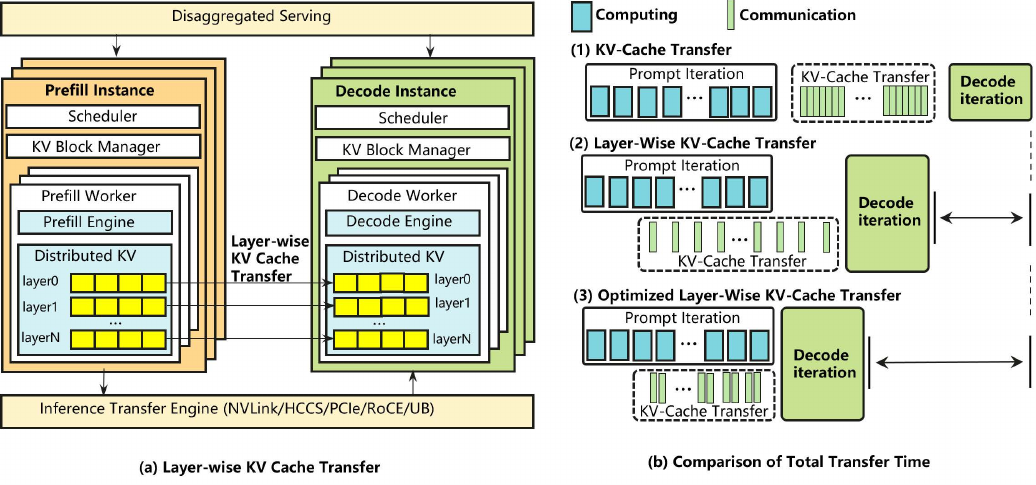}}
    \caption{P-D disaggregation with hierarchical KV transmission and pipelining. 
    Left: Hierarchically grouped KV transmission between the Prefill and Decode stages. 
    Right: Pipelined overlap of KV transmission and computation to hide communication latency.}
    \label{img:3_3}
    \end{center}
\end{figure}

\paragraph{Layer-wise KV Transmission:}
The KVCache is produced independently by each Transformer layer. 
\emph{EPD-Serve} partitions the cache into layer-wise transmission units, as shown in Figure \ref{img:3_3}, 
to reduce instantaneous bandwidth demand from one-shot KV transfer. 
When Prefill begins computation of the $(L+1)$-th layer, 
the KVCache from the $L$-th layer is asynchronously transmitted, achieving temporal overlap between communication and computation.

\paragraph{Hierarchically Grouped KV Transmission:}
KV transmission between Prefill and Decode typically requires a metadata handshake, which introduces unpredictable latency and delays hierarchical KV transmission, preventing full communication-computation overlap. 
To eliminate this misalignment, we propose a grouped transmission mechanism that packages KV data from multiple layers and schedules delayed transmission to align communication with model computation.

The mechanism consists of three core ideas:
\begin{itemize}
  \item Grouped Packaging: KVCache from adjacent layers is packaged and transmitted as a group, reducing communication frequency and handshake overhead. 
  The group size is dynamically determined based on MLP compute load and handshake latency.
  \item Asynchronous Overlap: An event-driven queue enables parallel execution of group transmission and MLP computation, maximizing the communication-computation overlap.
  \item Precise Scheduling: KV transmission is scheduled to avoid peak communication phases, reducing link contention and improving bandwidth efficiency.
\end{itemize}

Through the combined design of hierarchical KV transmission and grouped scheduling, \emph{EPD-Serve} achieves deep communication-computation overlap in the P-D stage. This significantly reduces cross-stage communication latency, mitigates bandwidth contention, and improves throughput and stability while maintaining low TTFT under high-concurrency multimodal workloads.

\subsection{Modality-Aware Multi-Path Scheduling Module}
\label{modality_multi_path}
In multimodal inference systems, different request types follow distinct execution paths: text-only requests require only Prefill and Decode, whereas multimodal requests additionally involve the Encode stage. A unified scheduling strategy causes cross-modal blocking and inefficient resource use. To address this, \emph{EPD-Serve} introduces a modality-aware multi-path scheduling module that selects inference paths based on request modality, achieving instance-level load balancing and maintaining high throughput and low latency under hybrid workloads.

\paragraph{Multi-Path Scheduling Strategy:}
Upon receiving a request from the API server, \emph{EPD-Serve} identifies whether the input contains non-text modalities, such as images, audio, or video. The request is then routed to the appropriate path: multimodal requests are processed through the E-P-D pipeline, while text-only requests follow the P-D pipeline. This routing mechanism creates separate execution pipelines for different modalities, preventing high-load multimodal requests from preempting resources required by text tasks and allowing each request type to execute efficiently on its designated path.

\paragraph{Instance-Level Dynamic Load Balancing:}
To further improve stability and utilization, \emph{EPD-Serve} employs a load-aware scheduling policy at the instance level. A global instance status table tracks metrics such as queue length, pending requests, and resource usage for each stage instance in real time. New requests are dispatched to the instance with the lowest load based on a least-loaded-first strategy, ensuring balanced distribution and preventing overload on individual nodes. Through the combination of modality-aware routing and instance-level load balancing, \emph{EPD-Serve} achieves heterogeneous traffic splitting at the request layer and dynamic load balancing at the execution layer, ensuring resource isolation and performance optimization for both multimodal and text tasks while maintaining scalability in high-concurrency scenarios.

\subsection{Logical Isolation and Physical Co-location Strategy}
\label{logical_physical}
To balance stage isolation with resource efficiency, \emph{EPD-Serve} adopts a combined design of logical isolation and physical co-location. At the logical layer, each stage is treated as an independent module, while at the physical layer, computing resources can be shared across stages. This approach enables targeted resource allocation and adaptive deployment of disaggregation layouts based on workload characteristics, modality distribution, and hardware availability, supporting dynamic selection among deployments such as E-P-D, EP-D, ED-P, E-PD, etc to optimize SLO outcomes.

\paragraph{Flexible Stage Disaggregation and Elastic Deployment:}
\emph{EPD-Serve} decouples the inference pipeline into independent instance processes for Encode, Prefill, and Decode stages, enabling separate scheduling and elastic scaling. A unified Proxy component performs cross-instance request routing and load balancing. Instances can be deployed on the same node or distributed across nodes depending on resource requirements, with low-overhead data transfer supported by asynchronous communication. Through software-level decoupling and communication optimization, \emph{EPD-Serve} can dynamically switch between deployment topologies and maintain scalability under heterogeneous resource environments.

\begin{figure}[htbp]
    \begin{center}
\centerline{\includegraphics[width=0.9\columnwidth]{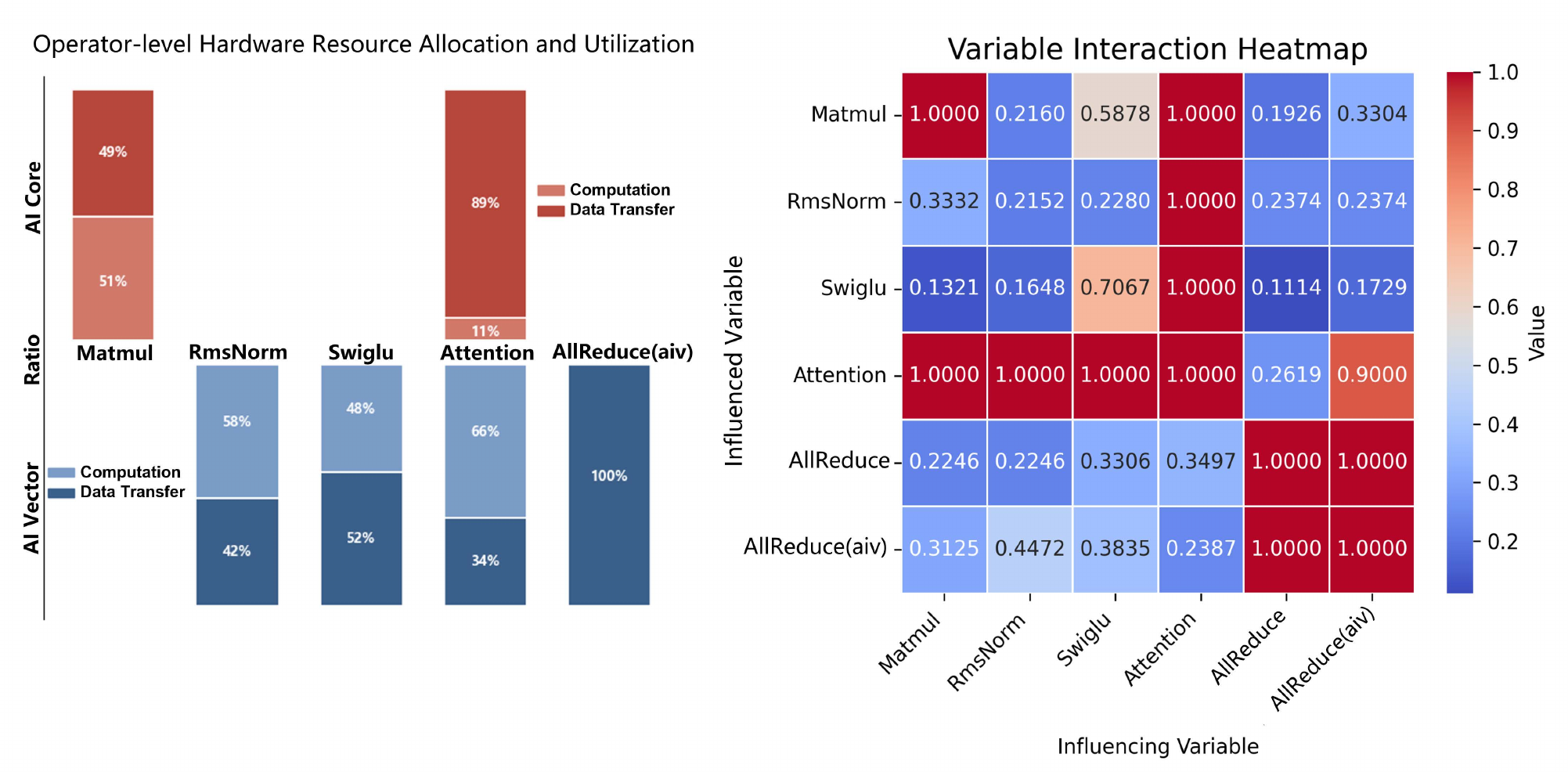}}
    \caption{Resource allocation and performance analysis of operator-level hardware co-location. 
The left subfigure illustrates the differences in hardware resource requirements, 
such as  AI Core, AI Vector, across various operators, as well as their relative proportions of computation versus data movement. 
The right subfigure presents a heatmap of latency increase during operator parallel execution, where smaller values indicate weaker performance degradation. 
Experimental results show that operators with significant differences in resource requirements exhibit minimal mutual interference when co-located, 
whereas operators with similar resource demands generate more pronounced performance interference under co-location.}
    \label{img:3_5}
    \end{center}
\end{figure}

\paragraph{Physical Co-Location and Spatial Multiplexing:}
While retaining independent scheduling at the software layer, \emph{EPD-Serve} supports physical co-location and spatial multiplexing of hardware resources. The Encode stage can be co-located with the Prefill or Decode stage on the same node, allowing complementary use of compute units and memory bandwidth. As shown in Figure \ref{img:3_5}, 
operators such as MatMul and AllReduce utilize different hardware components like AI Core and AI Vector. When one stage is waiting on communication, another stage can leverage idle compute cycles to execute its operators, achieving operator-level parallel reuse and improving throughput. With shared hardware resources enabled by co-location, the system increases utilization and throughput while maintaining predictable latency.

In summary, this strategy integrates software-level modularity with hardware-level resource sharing. The upper-layer architecture provides configurable disaggregation flexibility, while the lower-layer resources support spatial multiplexing, enabling a balanced trade-off between performance, energy efficiency, and latency across diverse multimodal workloads.

\section{Evaluation}
\label{inferification}
\subsection{Experiments Setup}
To evaluate the performance and scalability of \emph{EPD-Serve}, 
we conduct ablation studies and comparative experiments using diverse hybrid-modal datasets and several mainstream multimodal large models.

\paragraph{Datasets:}
Two representative datasets are selected for experimental evaluation, covering different multimodal inference scenarios:
VisualWebInstruct \cite{jia2025visualwebinstruct} is a text-image mixed instruction dataset. A subset of 512 samples is randomly selected for testing, consisting of 256 text-image requests and 256 text-only requests. 
All images are standardized to a resolution of 1280$\times$720, 
and text inputs contain 63.1 tokens on average. 
This dataset is used to evaluate the system's generalization performance in cross-modal inference scenarios.
ShareGPT-4o \cite{chen2025sharegpt} is a text-image dataset. A subset of 512 requests is randomly extracted, 
with an average image resolution of 802$\times$652 and an average text length of 9.6 tokens. 
This dataset is used to evaluate the system's performance on typical multimodal tasks.
Considering the dataset distribution and characteristics of multimodal understanding tasks, the output sequence length is uniformly fixed to 64 tokens in all experiments. 
Request injection is controlled using AISBench at 1-12 req/s to simulate different concurrency levels.
We record SLO attainment rate, throughput, TTFT, and TPOT as key metrics. 
For fair comparison across deployments, the per-NPU request rate is normalized by the number of NPUs in each deployment, ensuring a consistent single-device baseline.
Additionally, SLO differs by disaggregation strategy: when the Encode stage is disaggregated, $TTFT \leq 2000 ms$ and $TPOT \leq 80 ms$; when the Decode stage is disaggregated, $TTFT \leq 2000 ms$ and $TPOT \leq 50 ms$.

\paragraph{Models:}
We use two mainstream multimodal large models to evaluate performance consistency at different model scales: openPangu-7B-VL and Qwen3-VL-8B \cite{bai2025qwen3vltechnicalreport}.

\paragraph{Baseline and Deployment Notation:}
The baseline for performance comparison is the default monolithic architecture of vLLM v0.11.0, 
which executes the Encode, Prefill, and Decode sequentially on the same computing resource. 
In contrast, \emph{EPD-Serve} supports flexible disaggregation and co-location of the three stages. 
A unified notation is defined to characterize different deployments:
The symbol "-" denotes disaggregated deployment of distinct stages on separate hardware resources.
Parentheses "()" denote co-location of multiple stages on the same physical hardware, with logical isolation preserved.
For example, (E-PD) places Encode and the combined Prefill-Decode on the same device with logical isolation. EP-D deploys the Encode-Prefill and Decode stages on separate devices.

\paragraph{Hardware Platform:}
All experiments are conducted in a single-machine Ascend environment using the Ascend Atlas 800I A2 server with 64 GB of on-device memory per NPU. 
To ensure fairness and reproducibility, 
all comparative evaluations are conducted under identical hardware configurations.

\subsection{Effectiveness of EPD-Disaggregated Tensor Transmission}
We evaluate the two tensor transmission optimizations proposed in \emph{EPD-Serve} through ablation experiments on the ShareGPT-4o dataset, testing the E-P asynchronous feature prefetching mechanism and the P-D hierarchically grouped KV transmission mechanism separately.
Experiments are conducted under request rates of 2 and 3 req/s, and results are summarized in Table \ref{tab:4_1}.
When the E-P asynchronous feature prefetching mechanism is enabled, the system's TTFT decreased by approximately 16.6-21.7\% relative to the baseline.
This improvement mainly comes from preloading feature tensors, which overlaps E-P transmission with the Encode computation and effectively masks communication latency.
The P-D hierarchically grouped KV transmission mechanism yields an 11.9-16\% reduction in TTFT, showing that hierarchical grouping and delayed scheduling effectively reduce cross-instance KV transmission overhead.
When both mechanisms are enabled, TTFT decreases by 26.1-31.6\%,
demonstrating complementary latency-masking effects at different stages and jointly hiding end-to-end transmission costs.

\begin{table}[htbp]
    \centering
    \footnotesize
    \setlength{\tabcolsep}{4.5pt}
    \caption{Performance comparison of transmission optimizations in the E-P and P-D stages.}
    \small
    \begin{tabular}{ccccc}
    \toprule
    \multirow{2}{*}{Methods} & \multicolumn{2}{c}{Request Rate 2req/s} & \multicolumn{2}{c}{Request Rate 3req/s} \\ 
    \cline{2-5}\\[-8pt]
                         & TTFT (ms)            & TPOT(ms)      & TTFT(ms)             & TPOT(ms)      \\ \midrule
Baseline(E-P-D)          & 703.75               & 39.29         & 880.22               & 42.39         \\
w/ E-P Asynchronous Prefetching               & 586.87(-16.6\%)      & 38.36         & 688.86(-21.7\%)      & 41.5          \\
w/ P-D Hierarchically Grouped               & 590.80(-16.0\%)      & 39.42         & 775.83(-11.9\%)      & 43.89         \\
\emph{EPD-Serve}                & 481.38(-31.6\%)      & 38.20         & 650.51(-26.1\%)      & 43.95         \\
    \bottomrule
\end{tabular}
    \label{tab:4_1}
\end{table}

To further examine the performance gains of the E-P asynchronous feature prefetching and P-D hierarchically grouped KV transmission mechanisms, we conduct dedicated experiments and analyses as follows.

\subsubsection{Performance Analysis of E-P Stage Asynchronous Feature Prefetching}
To evaluate E-P transmission under different data scales, we use six groups of images with varying resolutions as input. We separately measure feature transmission latency and system scheduling latency to quantify the extent to which asynchronous transmission is hidden by system scheduling. The results are shown in Table \ref{tab:4_2}.

\begin{table}[t]
    \centering
    \footnotesize
    \setlength{\tabcolsep}{4.5pt}
    \caption{Performance of asynchronous feature prefetching in the E-P stage.}
    \small
    \begin{tabular}{ccccc}
    \toprule
    Image Resolution & Transmission Data & Transmission Latency(ms) & Scheduling Latency(ms) & Overlap Ratio \\\midrule
280$\times$280          & {[}100, 3584{]}   & 8.145                   & 30.803                & 100\%        \\
560$\times$560          & {[}400, 3584{]}   & 15.819                  & 42.406                & 100\%        \\
640$\times$960          & {[}529, 3584{]}   & 17.019                  & 49.549                & 100\%        \\
720$\times$1280         & {[}1196, 3584{]}  & 38.776                  & 81.028                & 100\%        \\
1080$\times$1920        & {[}2691, 3584{]}  & 80.771                  & 151.77                & 100\%        \\
4096$\times$3112        & {[}16206, 3584{]}  & 729.724                 & 728.109               & 99.78\%     \\\bottomrule
\end{tabular}
    \label{tab:4_2}
\end{table}

\begin{figure}[h]
  \centering
  \subfigure[Profiling of layer-wise KV transmission with a 1024-sequence input, where the KV transmission overlap ratio is only 15.27\%.]{
  \label{img:4_1:a}
  \includegraphics[width=0.9\linewidth]{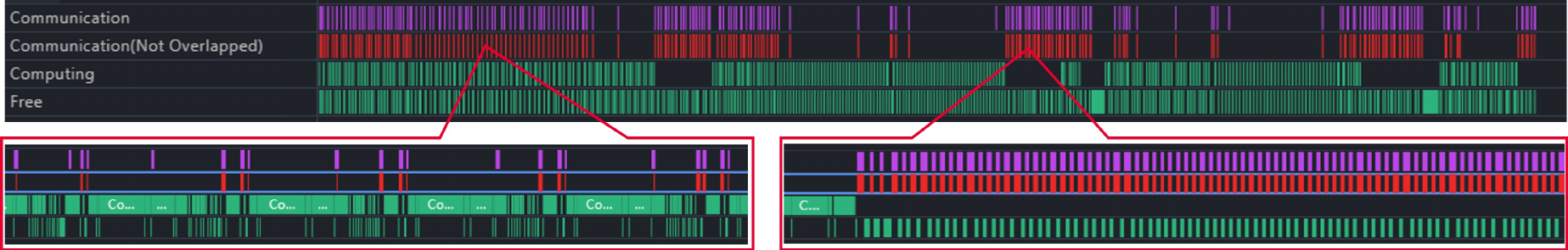}}
  \\
  \subfigure[Profiling of hierarchically grouped KV transmission with a 1024-sequence input, where the KV transmission overlap ratio reaches 98.78\%.]{
  \label{img:4_1:b}
  \includegraphics[width=0.9\linewidth]{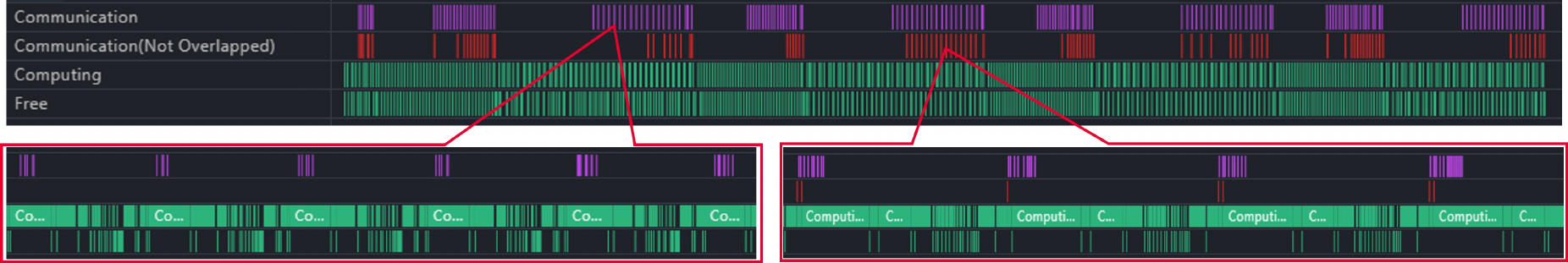}}
  \\
  \subfigure[Profiling of layer-wise KV transmission with a 2048-sequence input, where the KV transmission overlap ratio is only 25.08\%.]{
  \label{img:4_1:c}
  \includegraphics[width=0.9\linewidth]{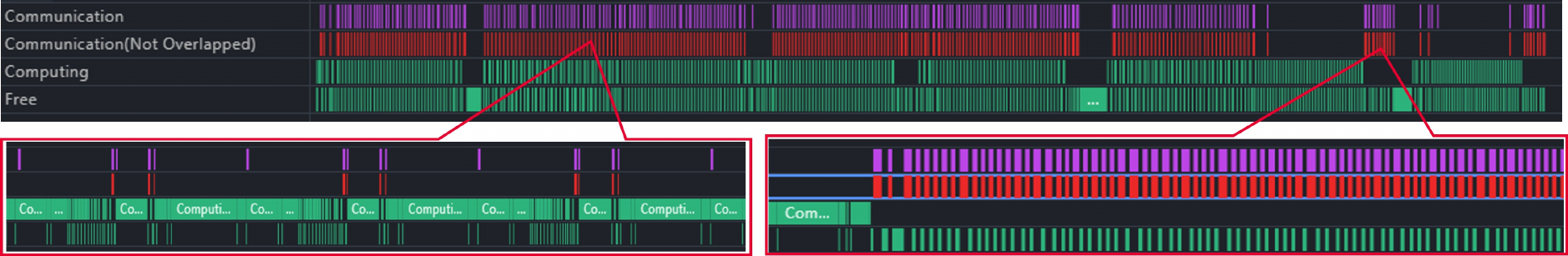}}
  \\
  \subfigure[Profiling of hierarchically grouped KV transmission with a 2048-sequence input, where the KV transmission overlap ratio reaches 99.92\%.]{
  \label{img:4_1:d}
  \includegraphics[width=0.9\linewidth]{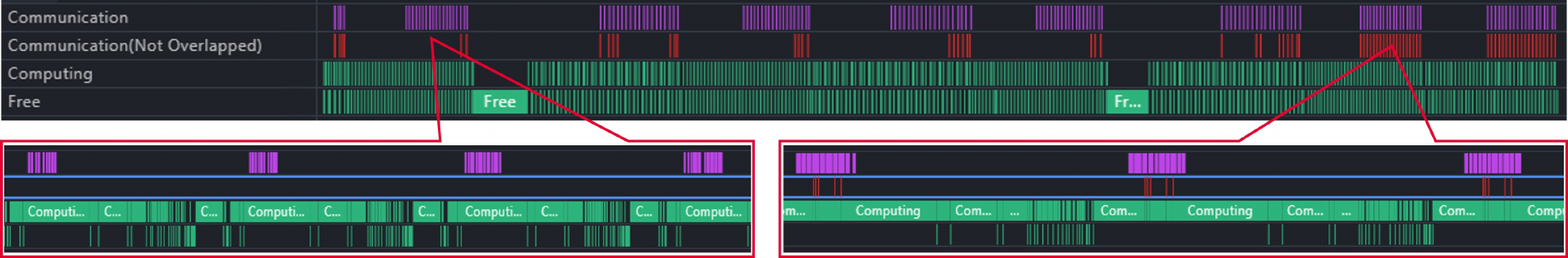}}
  \caption{Performance comparison of layer-wise KV transmission and optimized hierarchically grouped KV transmission under different input sequence lengths.}
  \label{img:4_1}
\end{figure}

The results demonstrate that feature transmission can be effectively overlapped by both inter-instance and intra-instance scheduling.
When the input image resolution is below 4K, 
the feature transmission latency remains consistently lower than the scheduling latency, 
enabling nearly 100\% transmission overlap ratio for the E-P stage. 
As image resolution increases, 
feature transmission latency increases rapidly and exceeds scheduling latency, 
resulting in only partial masking of the transmission process. 
For instance, at 4K resolution, the overlap ratio drops to 99.78\%.
In conclusion, the E-P stage asynchronous feature prefetching mechanism achieves nearly complete overlap efficiency for image inputs at mainstream resolutions, 
with a minor degradation in overlap capability under higher-resolution scenarios.

\subsubsection{Performance Analysis of Hierarchically Grouped KV Transmission}
To evaluate KV transmission efficiency under different data volumes, we compare the performance of the layer-wise transmission before and after applying hierarchically grouped optimization at input sequence lengths of 1024 and 2048 tokens, with a concurrency of 16. The results are shown in Figure \ref{img:4_1}.

The results show that hierarchically grouped KV transmission delivers substantial performance gains at both sequence lengths.
With input lengths of 1024 and 2048 tokens, the KV transmission overlap ratio increases by 83.51\% and 74.84\% over the baseline, respectively.
In general, overlap improves as sequence length grows.
For example, the overlap ratio of the baseline rises from 15.27\% at a sequence length of 1024 tokens to 25.08\% at 2048 tokens. 
This is because longer sequences yield more computation in the Prefill stage, reducing the exposed KV transmission time.

In contrast, the optimized scheme already achieves efficient overlap through computation-communication alignment, so its overlap ratio is less sensitive to sequence length. 
Furthermore, hierarchically grouped packaging combined with the asynchronous transmission interface of Mooncake increases the payload size of each transfer and improves bandwidth utilization. 
Compared with baseline layer-wise transmission, average bandwidth utilization increases by 58\% at a sequence length of 1024 tokens and 10\% at 2048 tokens, as shown in Table \ref{tab:4_3}.  
The improvement is more pronounced for small inputs, where baseline layer-wise transmission results in small KV payloads per transfer, while grouped packaging increases transfer granularity and yields substantially higher bandwidth efficiency.

\begin{table}[t]
    \centering
    \footnotesize
    \setlength{\tabcolsep}{1.5pt}
    \caption{Performance comparison of layer-wise KV transmission before and after optimization.}
    \small
    \begin{tabular}{ccccccc}
    \toprule
Input Length          & Method    & KV Latency(ms) & Exposed Latency(ms) & Prefill Latency (ms) & Overlap Ratio & Bandwidth (GB/s) \\\midrule
\multirow{2}{*}{1024} & Baseline  & 1127.45        & 955.24              & 6793.50              & 15.27\%      & 7.98             \\\cline{2-7}\\[-8pt]
                      & Optimized & 715.53         & 8.76                & 6610.57              & 98.78\%      & 12.58            \\\midrule
\multirow{2}{*}{2048} & Baseline  & 1688.40        & 1264.87             & 14349.47             & 25.08\%      & 10.66            \\\cline{2-7}\\[-8pt]
                      & Optimized & 1536.49        & 1.16                & 14261.21             & 99.92\%      & 11.71            \\\bottomrule   
\end{tabular}
    \label{tab:4_3}
\end{table}

\subsection{Benefits of Encode Disaggregation}

\begin{figure}[!h]
    \centering
    \subfigure{
    \includegraphics[width=0.4\textwidth]{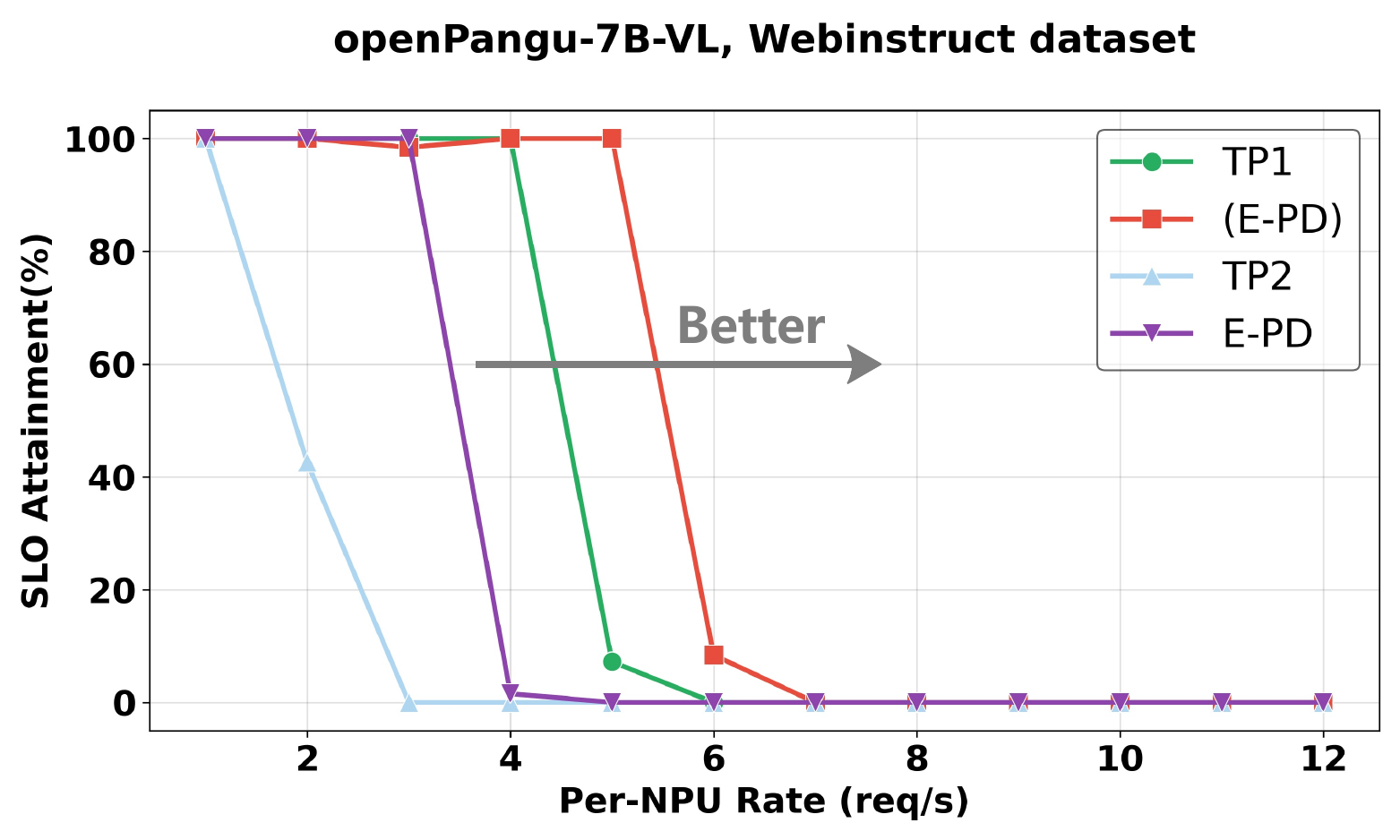}
    \hspace{0.1in}
    \includegraphics[width=0.4\textwidth]{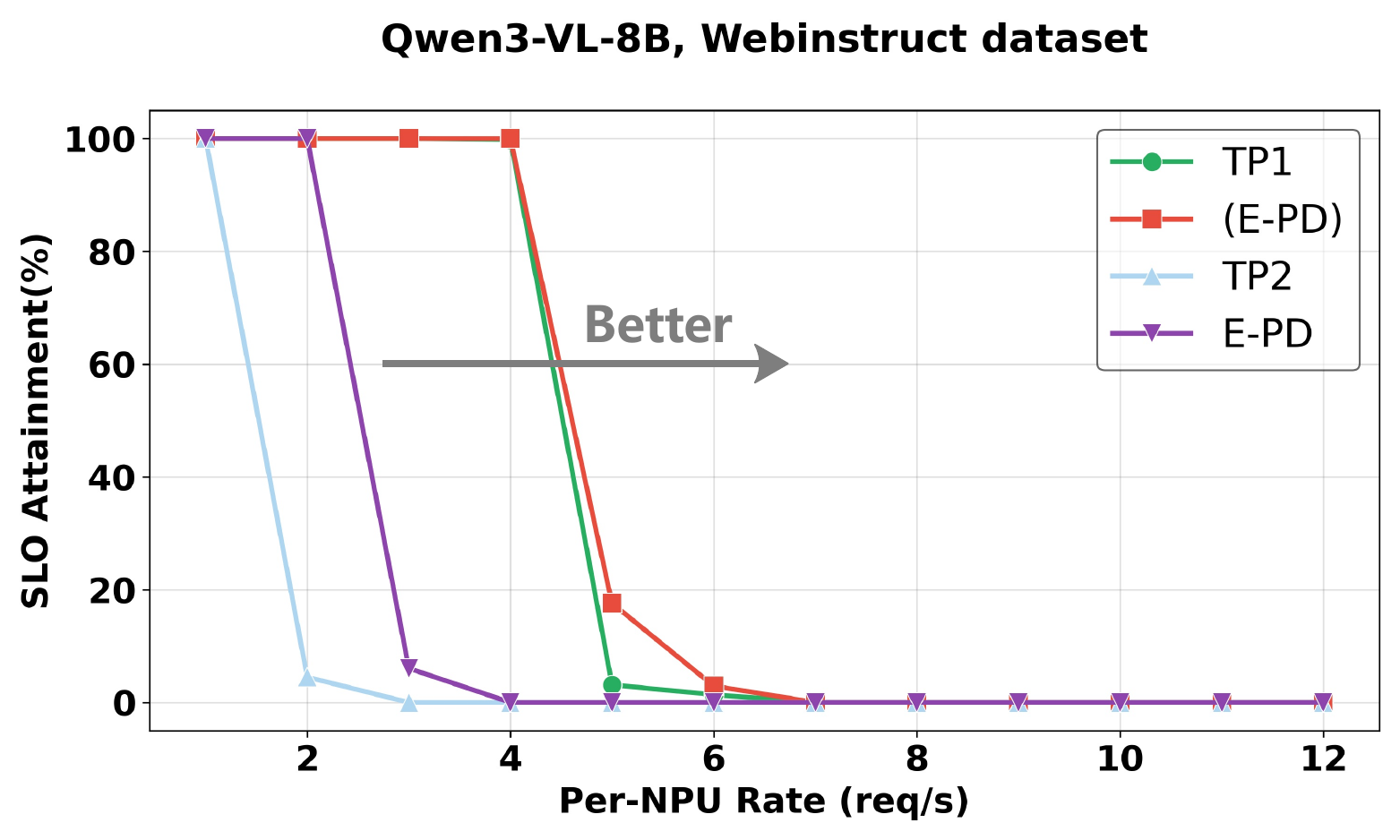}
    }
    \par\noindent\setlength{\parskip}{0pt}  
    \vspace{-5pt}  
    \subfigure{
    \includegraphics[width=0.4\textwidth]{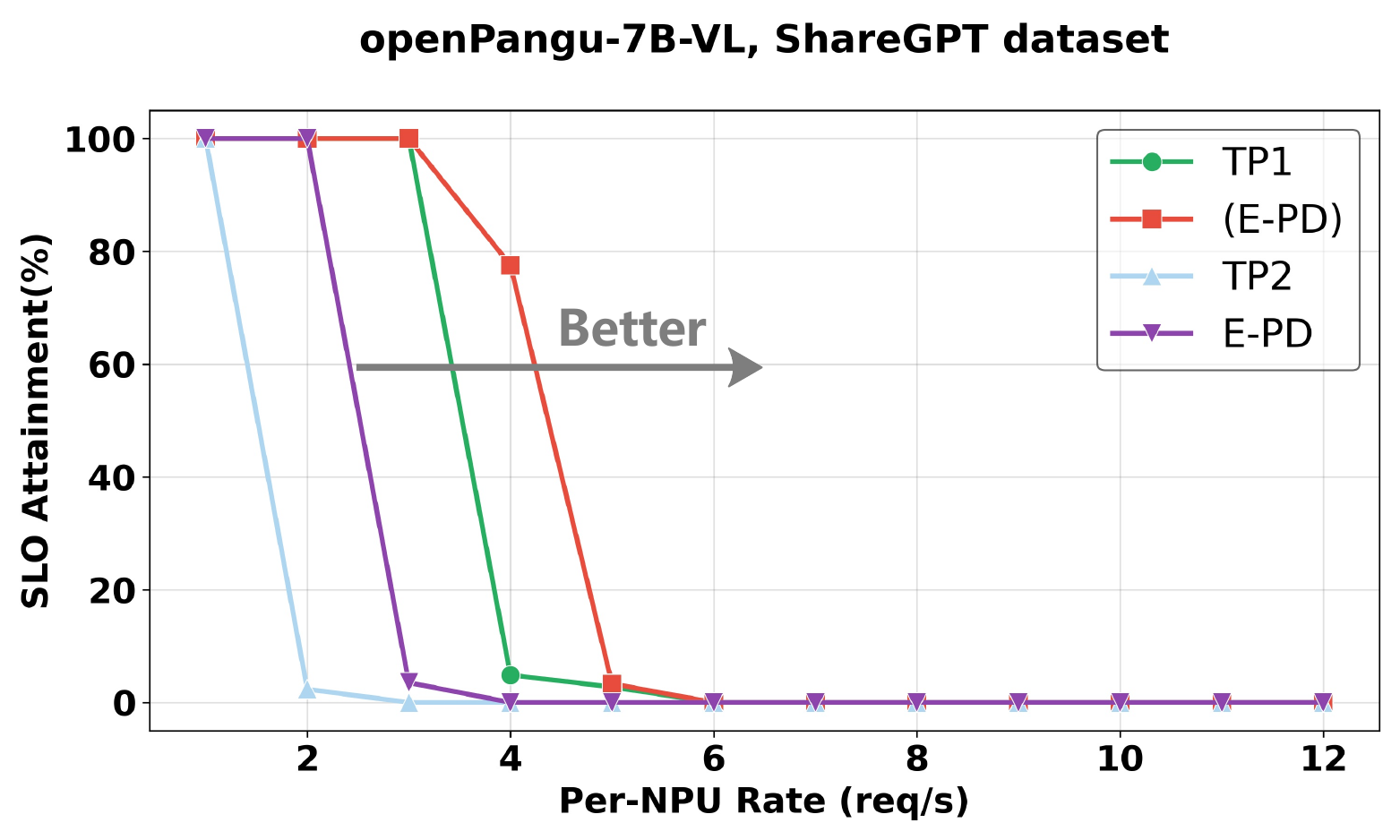}
    \hspace{0.1in}
    \includegraphics[width=0.4\textwidth]{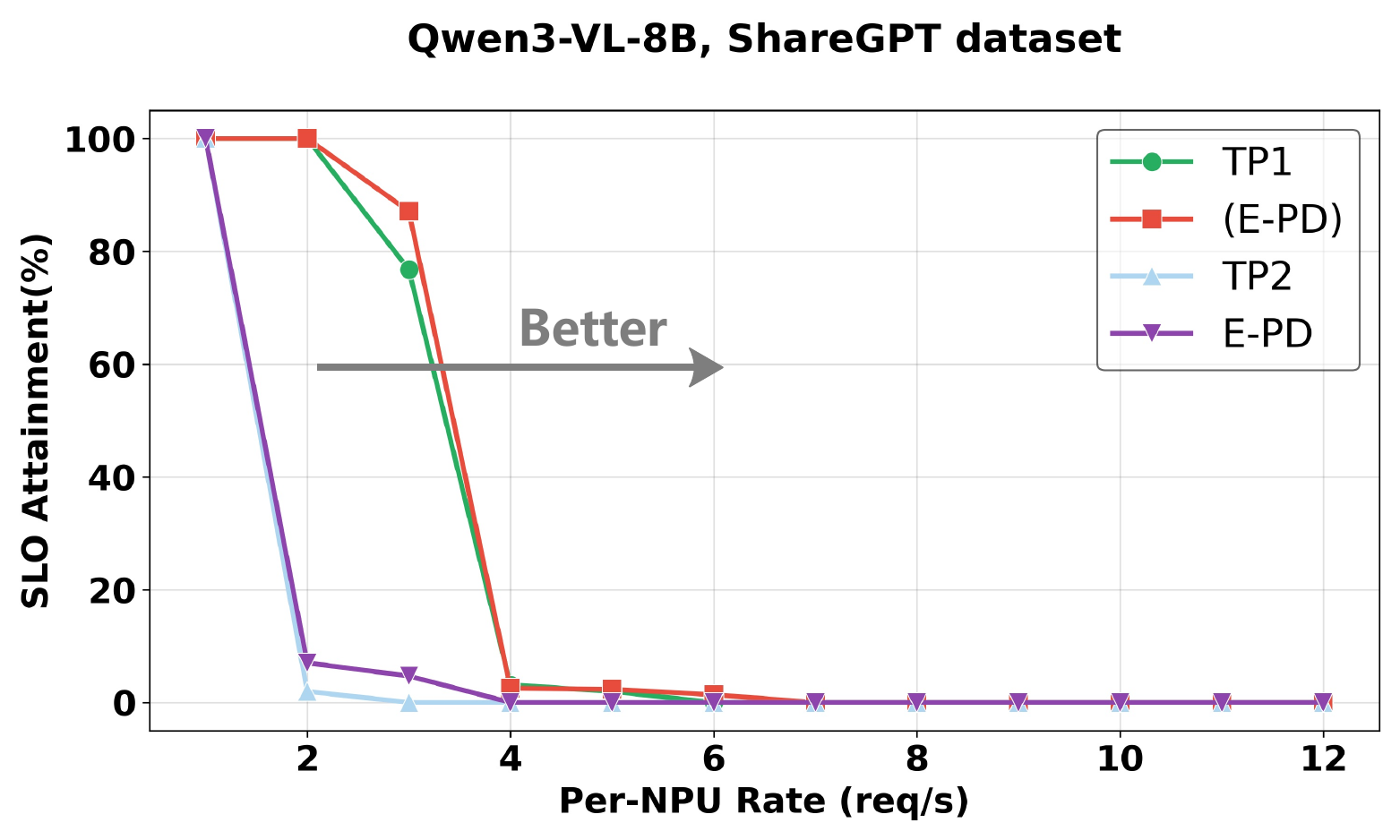}
    }
    \caption{Comparison of SLO attainment rate between Encode-stage disaggregated and monolithic deployments. 
    The (E-PD) deployment, which disaggregates and co-locates the E and PD on a single NPU, 
    achieves a higher SLO attainment rate than the TP1 baseline.}
    \label{img:4_2:a}
\end{figure}

\begin{figure}[!h]
    \centering
    \subfigure{
    \includegraphics[width=0.4\textwidth]{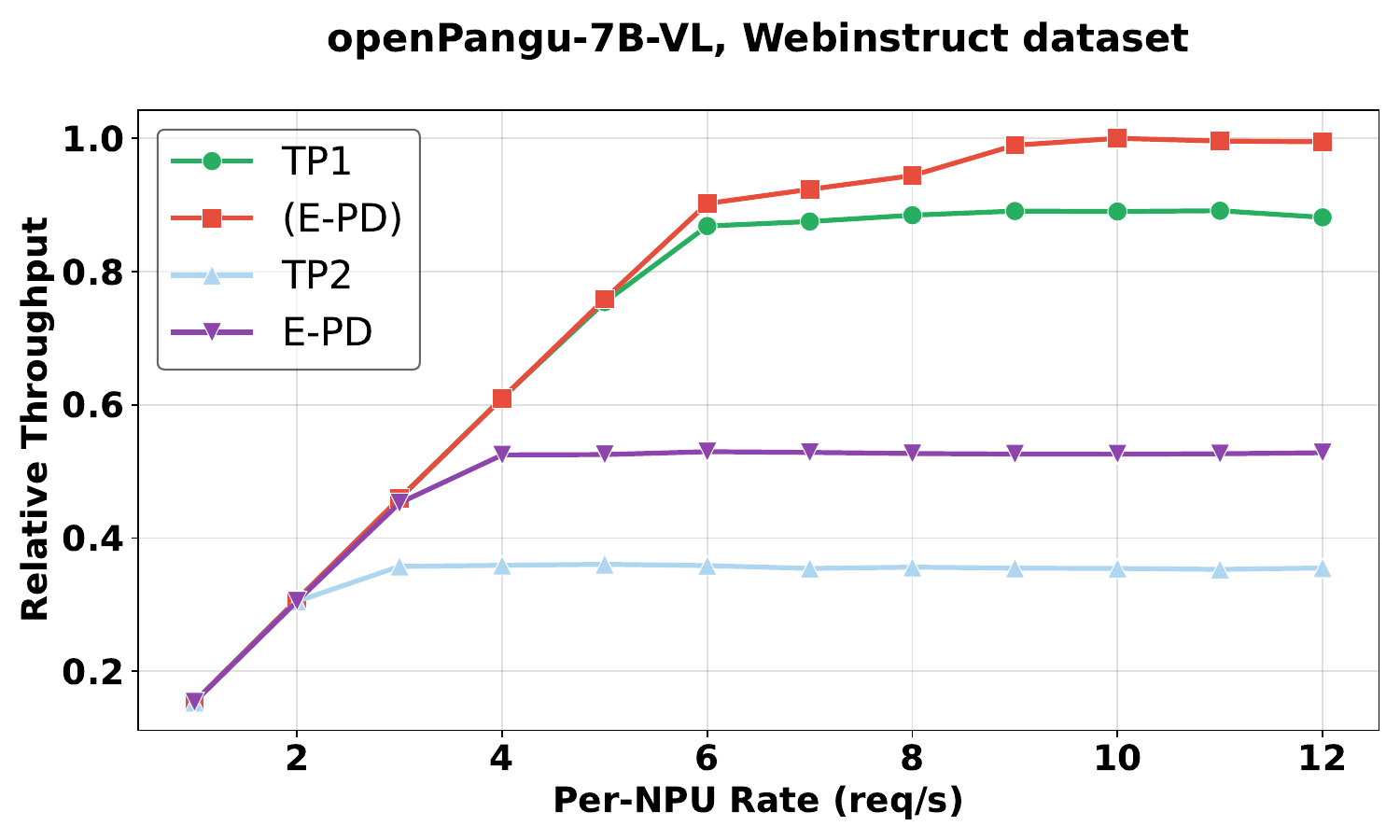}
    \hspace{0.1in}
    \includegraphics[width=0.4\textwidth]{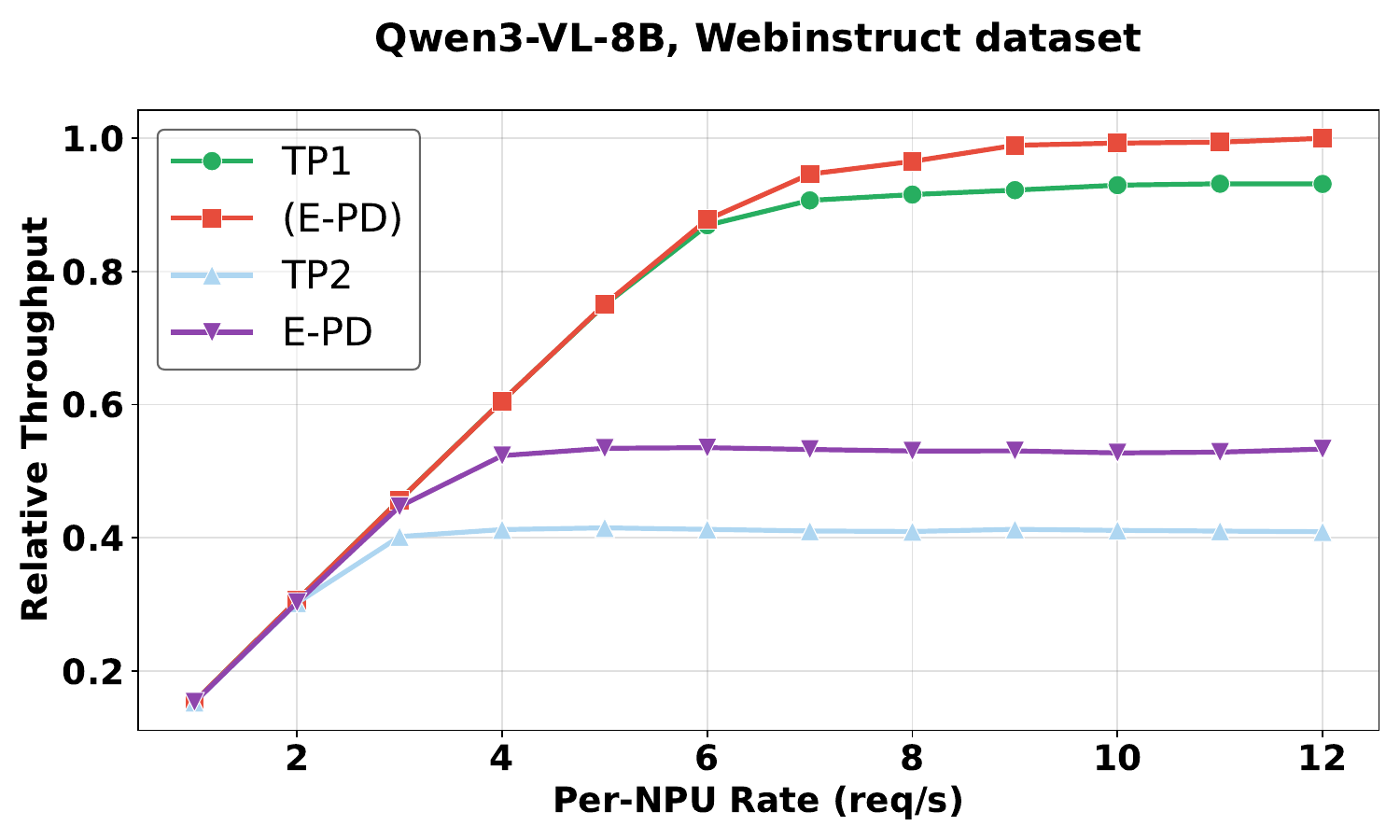}
    }
    \par\noindent\setlength{\parskip}{0pt}  
    \vspace{-5pt}  
    \subfigure{
    \includegraphics[width=0.4\textwidth]{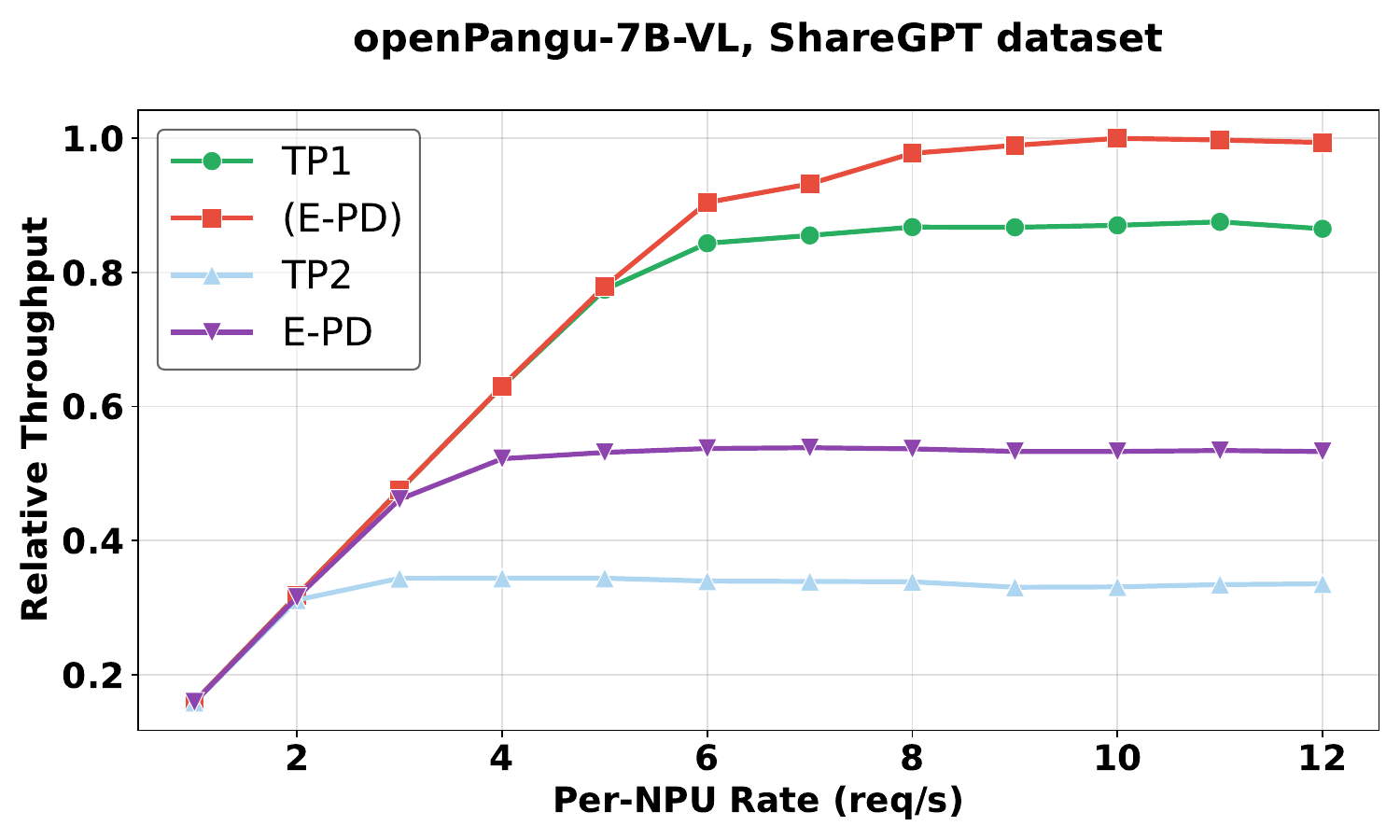}
    \hspace{0.1in}
    \includegraphics[width=0.4\textwidth]{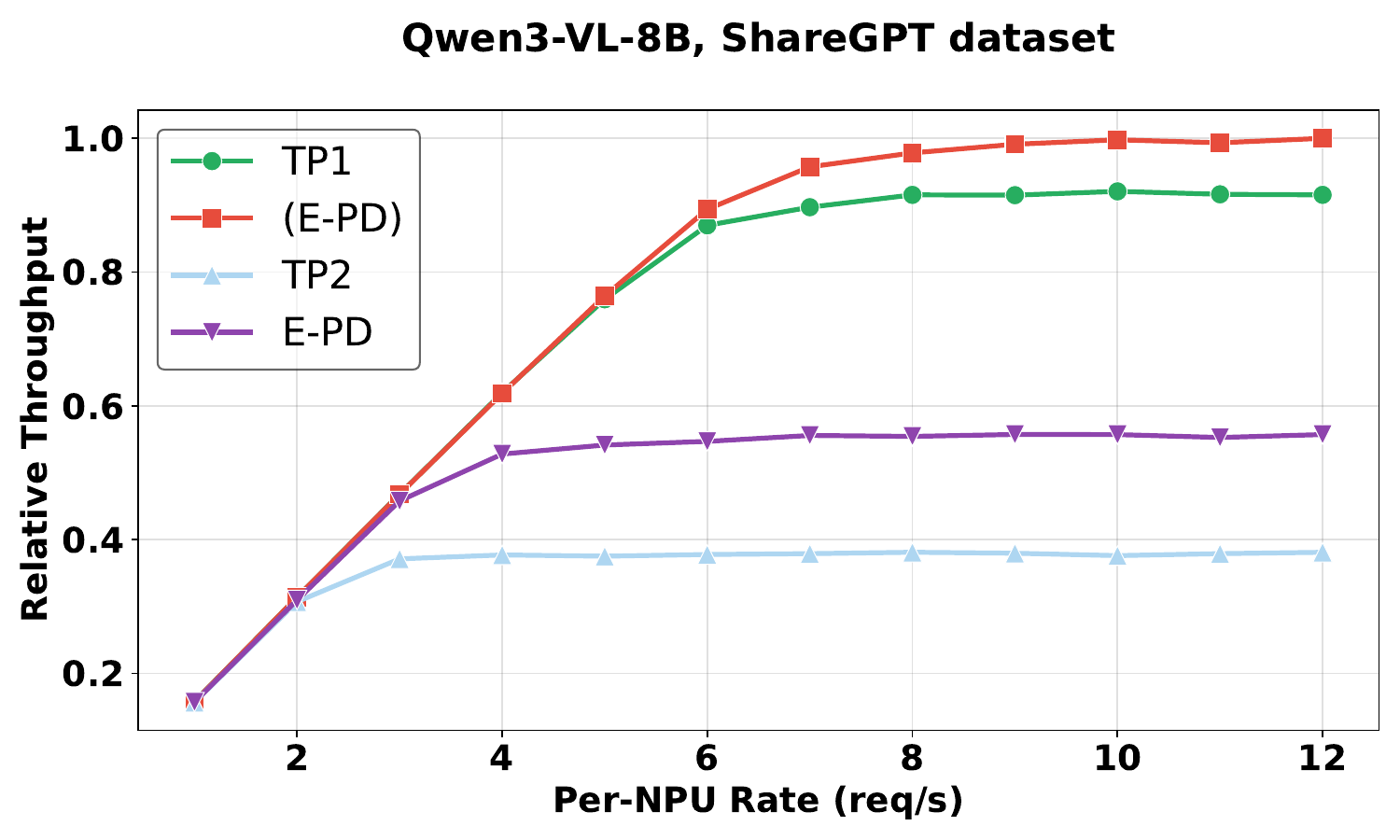}
    }
    \caption{Comparison of throughput performance between Encode-stage disaggregated and monolithic deployments. 
    The (E-PD) deployment, which disaggregates and co-locates the E and PD on a single NPU, 
    outperforms the TP1 baseline in throughput under single-NPU settings.}
    \label{img:4_2:b}
\end{figure}

\begin{figure}[!h]
    \centering
    \subfigure{
    \includegraphics[width=0.4\textwidth]{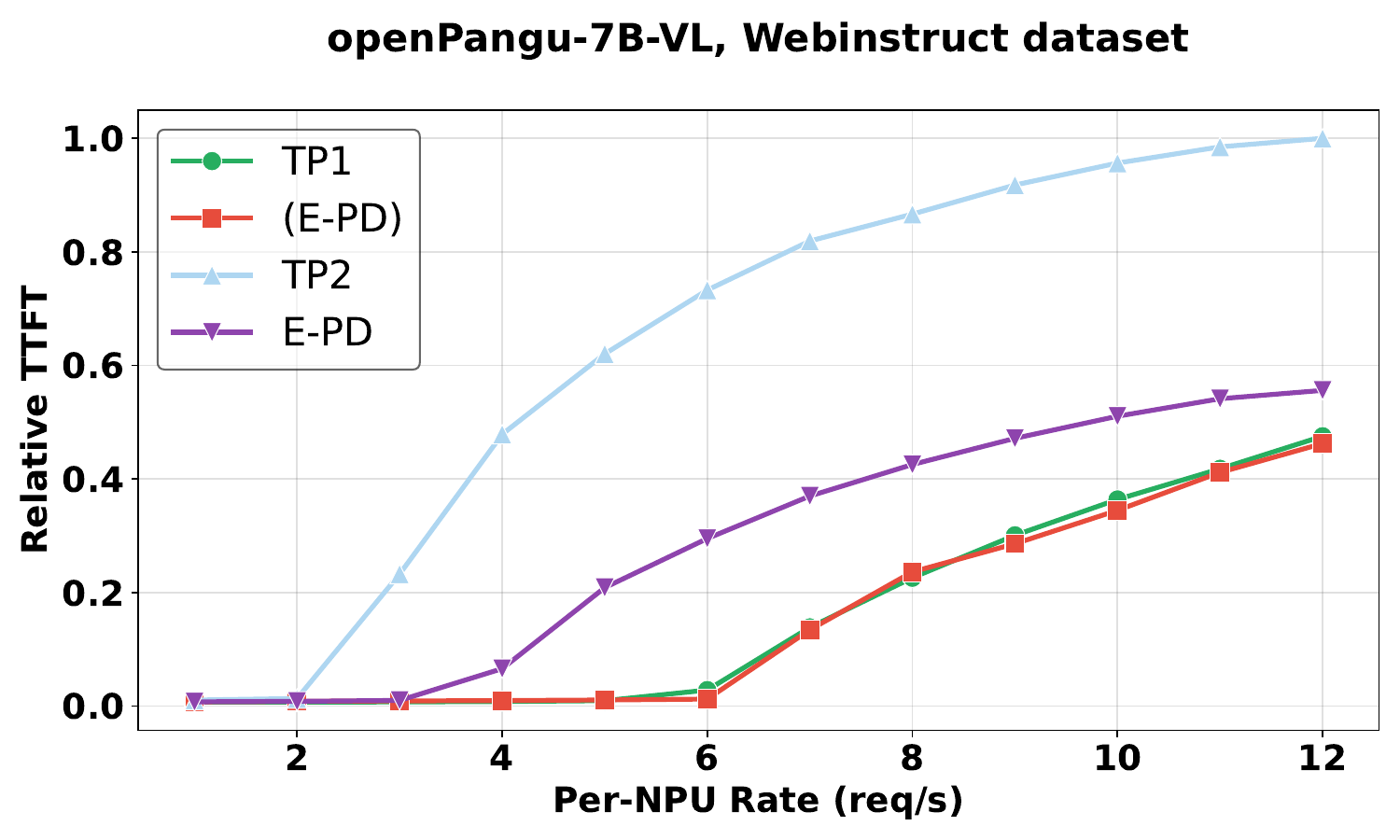}
    \hspace{0.1in}
    \includegraphics[width=0.4\textwidth]{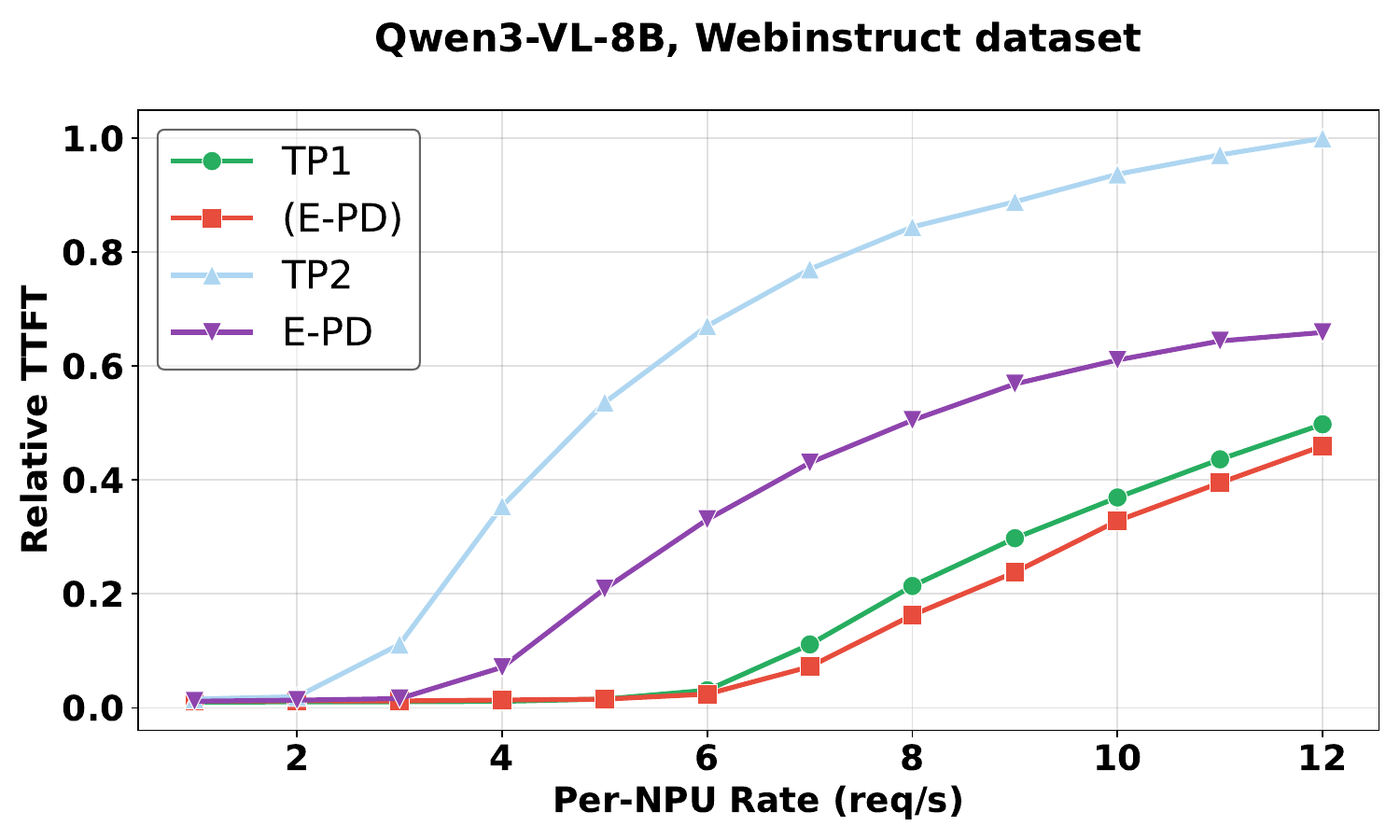}
    }
    \par\noindent\setlength{\parskip}{0pt}  
    \vspace{-5pt}  
    \subfigure{
    \includegraphics[width=0.4\textwidth]{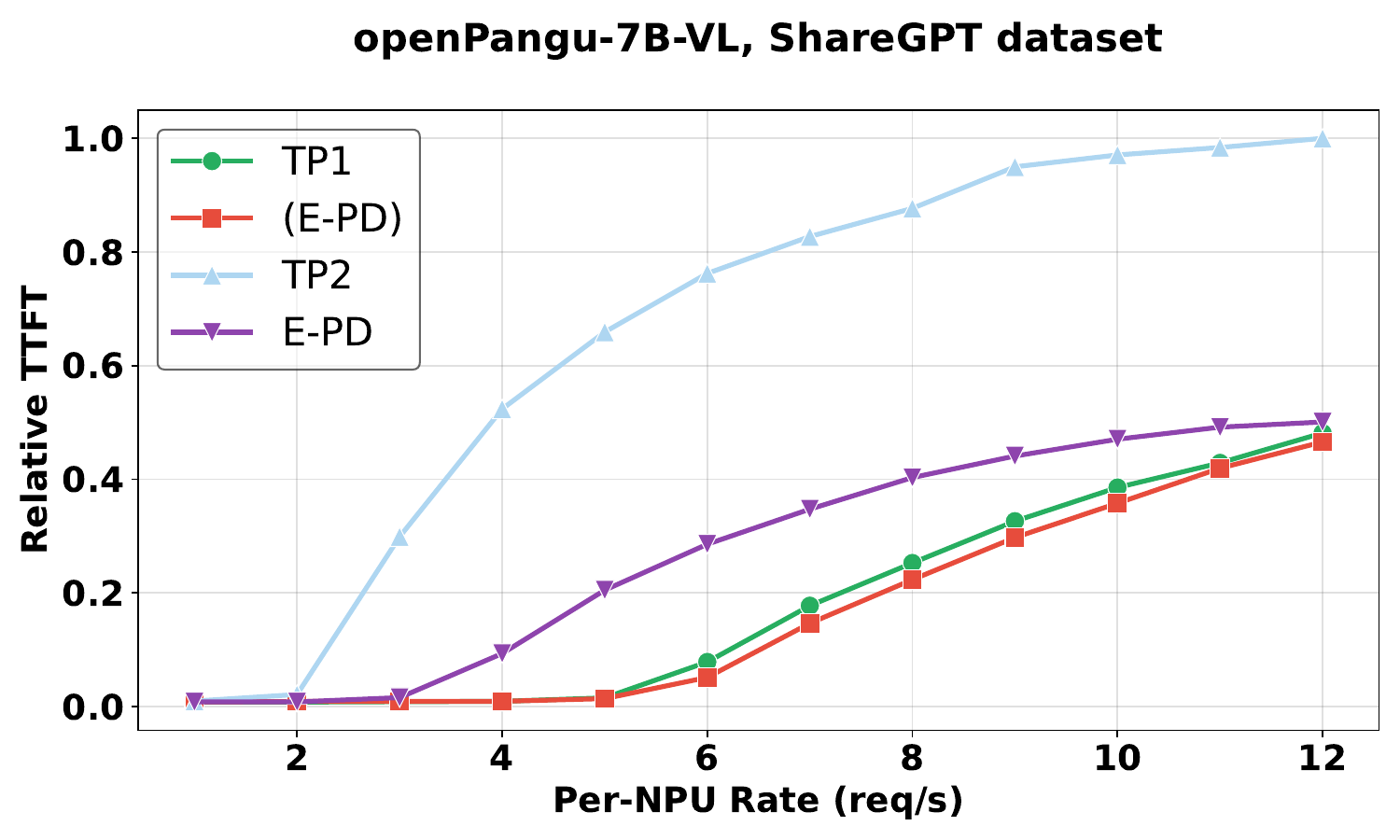}
    \hspace{0.1in}
    \includegraphics[width=0.4\textwidth]{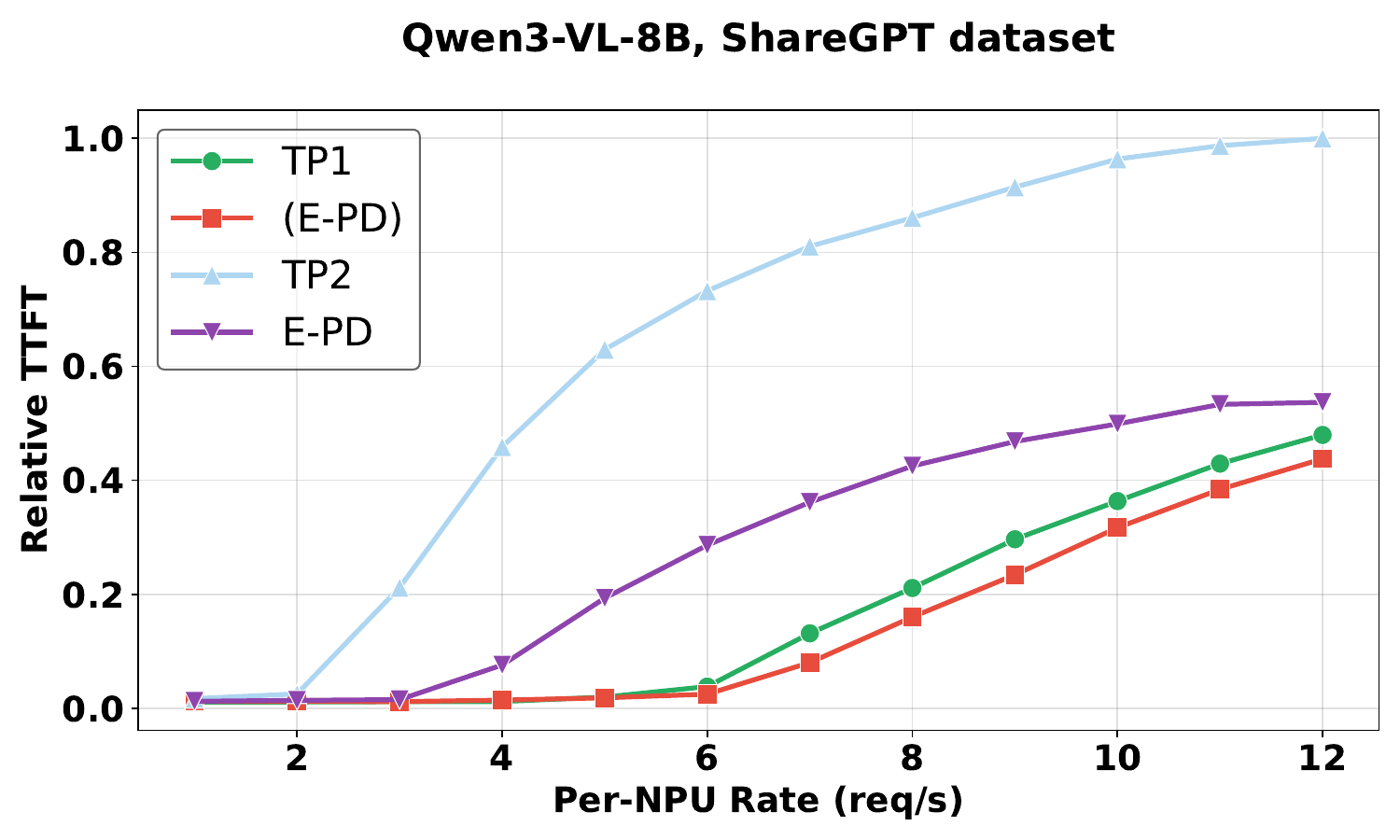}
    }
    \caption{Comparison of TTFT latency between Encode-stage disaggregated and monolithic deployments. 
   The (E-PD) deployment, which disaggregates and co-locates the E and PD on a single NPU, 
   delivers lower TTFT latency than the TP1 baseline under single-NPU settings.}
   \label{img:4_2:c}
\end{figure}

\begin{figure}[!h]
    \centering
    \subfigure{
    \includegraphics[width=0.4\textwidth]{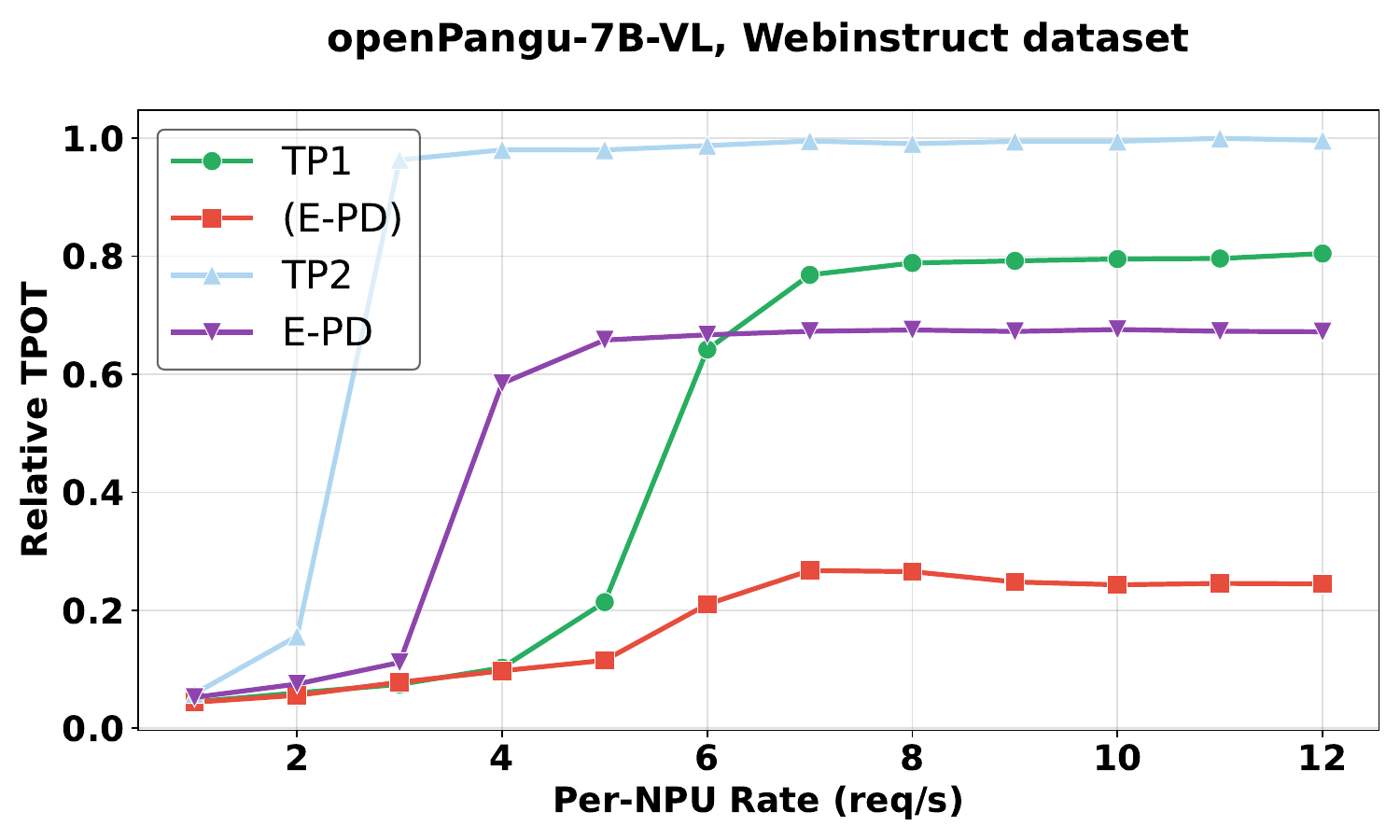}
    \hspace{0.1in}
    \includegraphics[width=0.4\textwidth]{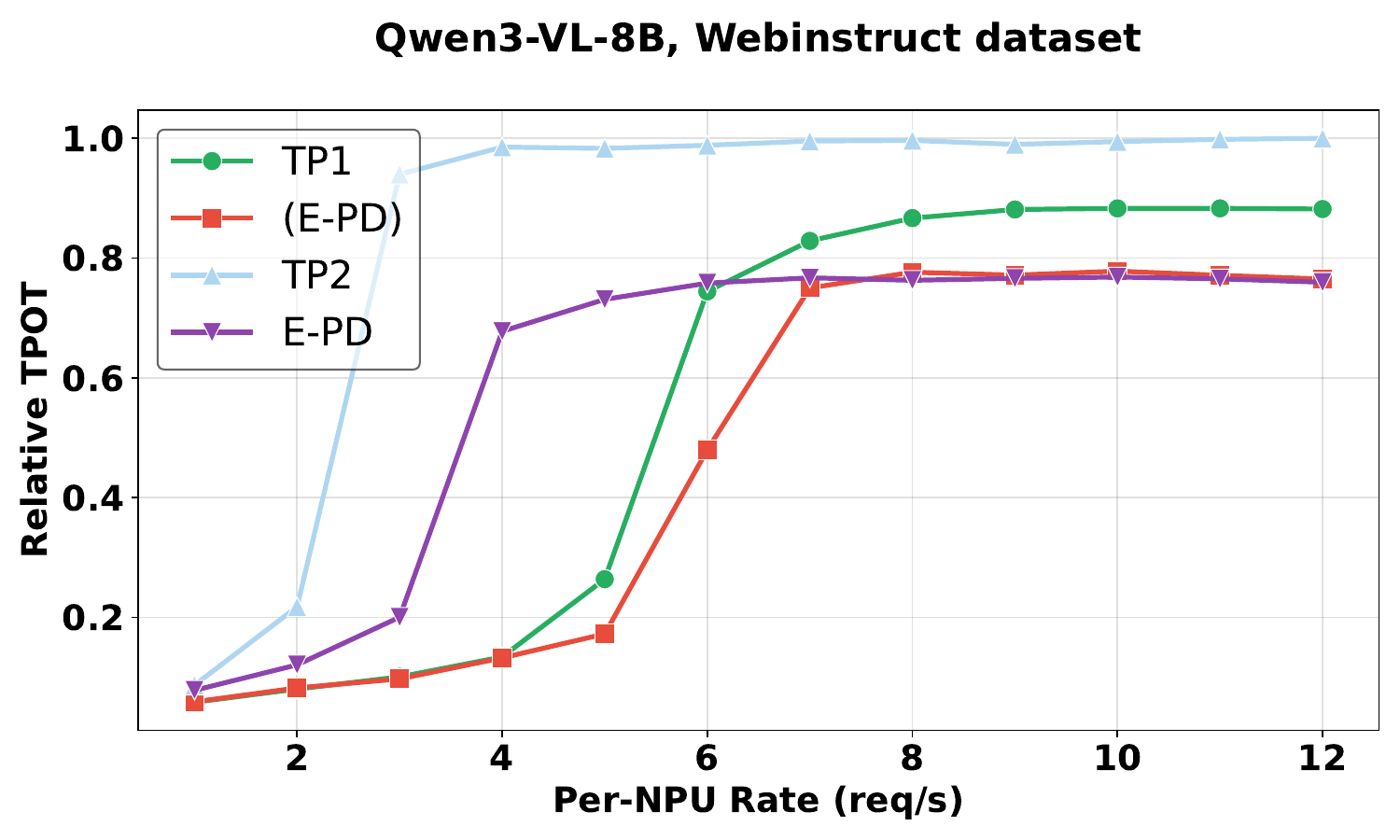}
    }
    \par\noindent\setlength{\parskip}{0pt}  
    \vspace{-5pt}  
    \subfigure{
    \includegraphics[width=0.4\textwidth]{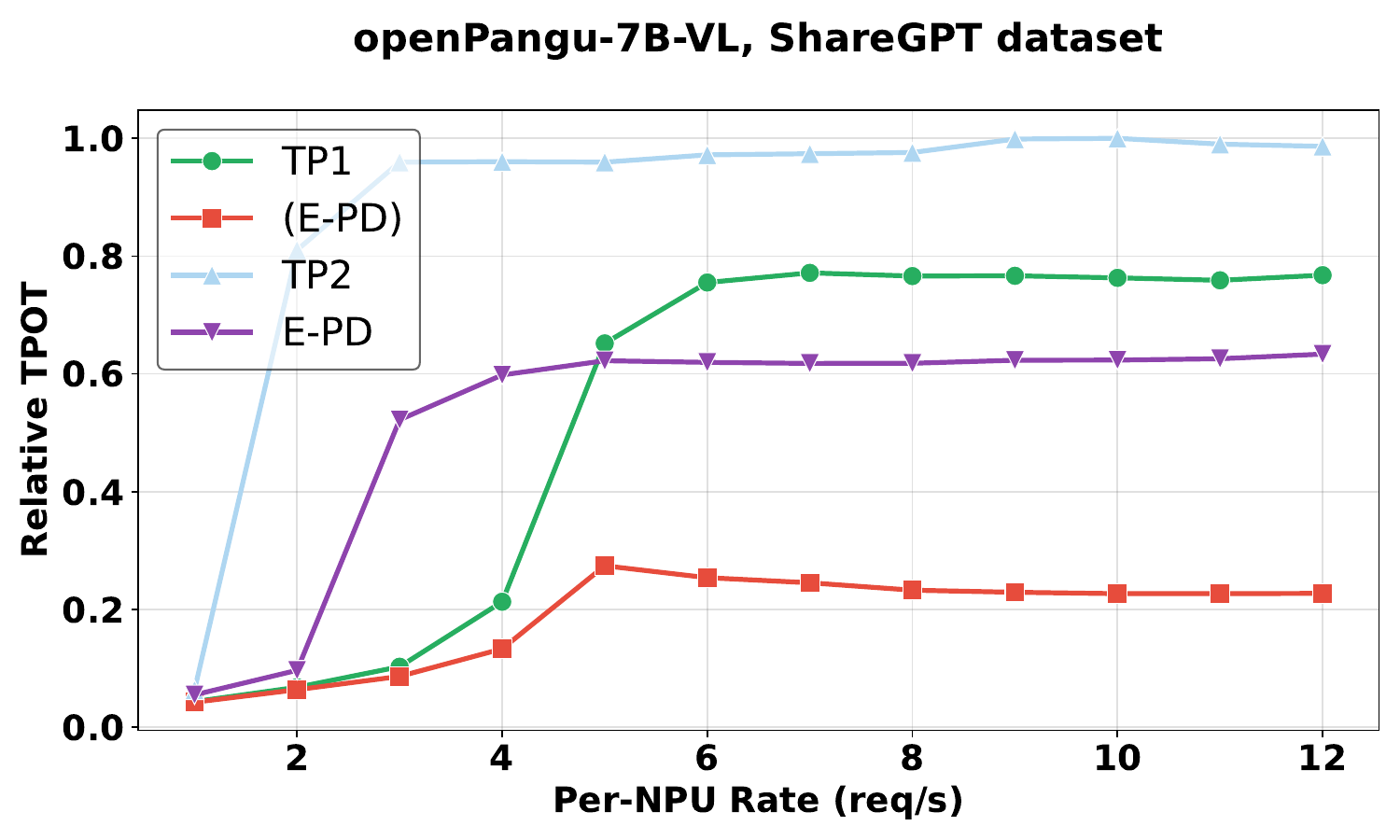}
    \hspace{0.1in}
    \includegraphics[width=0.4\textwidth]{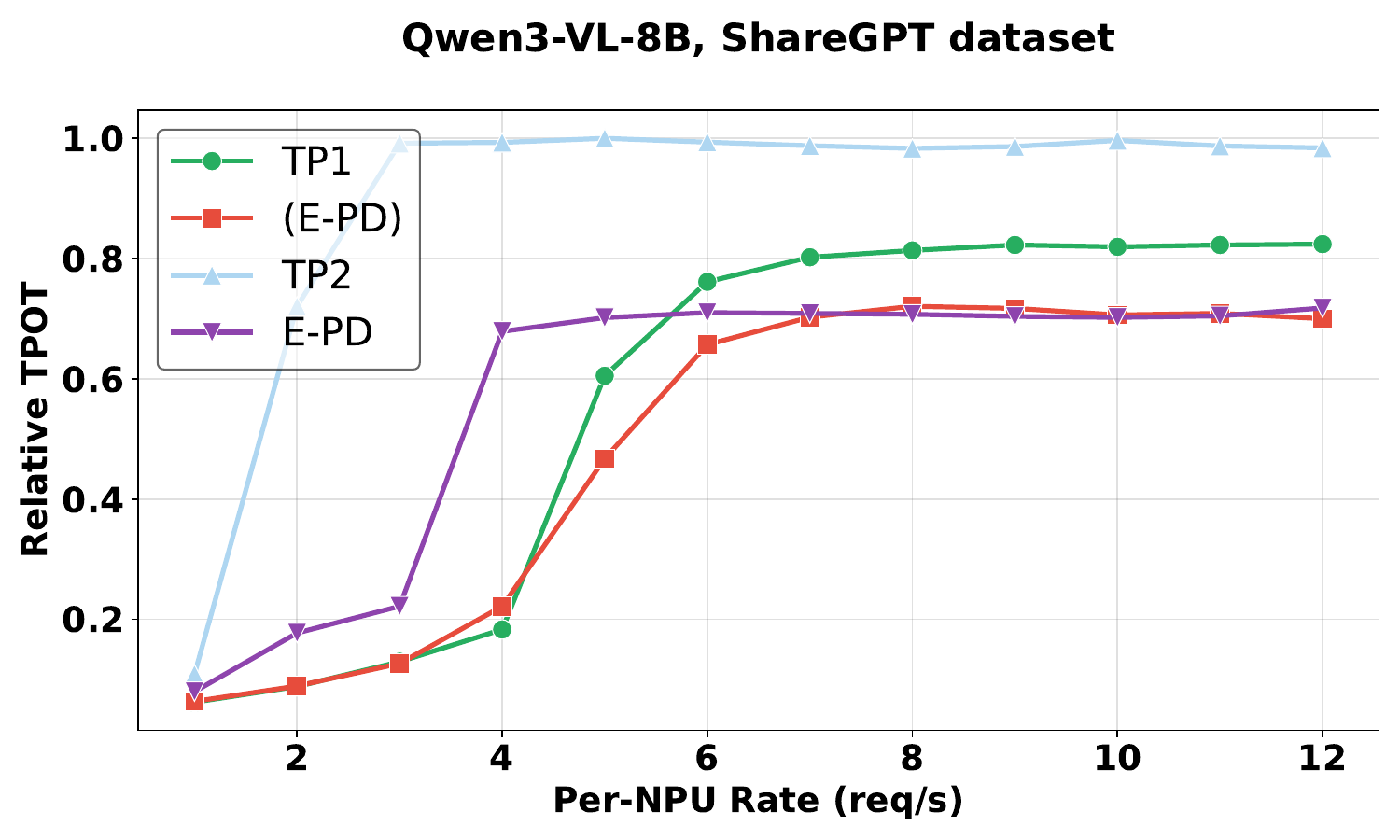}
    }
    \caption{Comparison of TPOT latency between Encode-stage disaggregated and monolithic deployments. 
   The (E-PD) deployment, which disaggregates and co-locates the E and PD on a single NPU, 
   still maintains lower TPOT latency over the TP1 baseline under single-NPU settings.}
   \label{img:4_2:d}
\end{figure}

Given the complex impact of stage disaggregation on multimodal inference performance, we first analyze the effects of disaggregating the Encode stage. 
This section evaluates the performance differences of multiple deployments, including TP1, TP2, (E-PD), and E-PD, derived from the Encode disaggregation. 
\cref{img:4_2:a,img:4_2:b,img:4_2:c,img:4_2:d} illustrate the trends of four key metrics, SLO attainment rate, system throughput, TTFT, and TPOT, under varying request rates on the VisualWebInstruct and ShareGPT-4o datasets.

The results show that the E-PD deployment places the Encode stage and the LLM inference stages on separate NPUs. Because the Encode stage has relatively low computational demand, dedicating an independent NPU to it leads to poor hardware utilization and inferior performance compared with the TP1 baseline across all metrics, including SLO attainment rate, throughput, TTFT, and TPOT.
In contrast, the (E-PD) deployment leverages physical co-location, allowing the Encode and PD stages to complement each other in compute and memory usage. This enables more effective reuse of AI Core and memory resources and yields consistently superior performance across models and datasets. 
For openPangu-7B-VL, compared with the TP1 baseline, (E-PD) maintains higher SLO attainment under all loads and, at 12 req/s, improves throughput by 12.87-14.88\%, reduces TTFT by 2.7-3.25\%, and reduces TPOT by 69.58-70.39\%. For Qwen3-VL-8B, (E-PD) similarly outperforms TP1, with throughput gains of 7.36-9.24\%, TTFT reductions of 7.65-8.74\%, and TPOT reductions of 13.25-15.05\%.

Further analysis of TTFT shows that (E-PD) surpasses TP1 once the load exceeds 6 req/s and provides more than a 2.7\% improvement at 12 req/s. This indicates that under high concurrency, disaggregating the Encode stage enables more efficient multi-core utilization on the shared NPU, thereby improving end-to-end inference latency. For TPOT, although Decode dominates token-generation latency, Encode disaggregation shortens inference delay when Decode is momentarily blocked by new requests, yielding better TPOT performance than TP1. By contrast, TP2 increases parallelism through additional NPUs, but its inter-NPU synchronization overhead severely degrades performance, making it the worst-performing deployment.
In summary, (E-PD) achieves both logical disaggregation and physical co-location of E and PD stages. This enables PD idle compute windows to be effectively reclaimed by the Encode stage, improving throughput while maintaining SLO stability and significantly reducing TTFT and TPOT. These results validate the advantages of Encode-stage disaggregation combined with physical co-location.

\subsection{Benefits of Decode Disaggregation}
\begin{figure}[!h]
    \centering
    \subfigure{
    \includegraphics[width=0.4\textwidth]{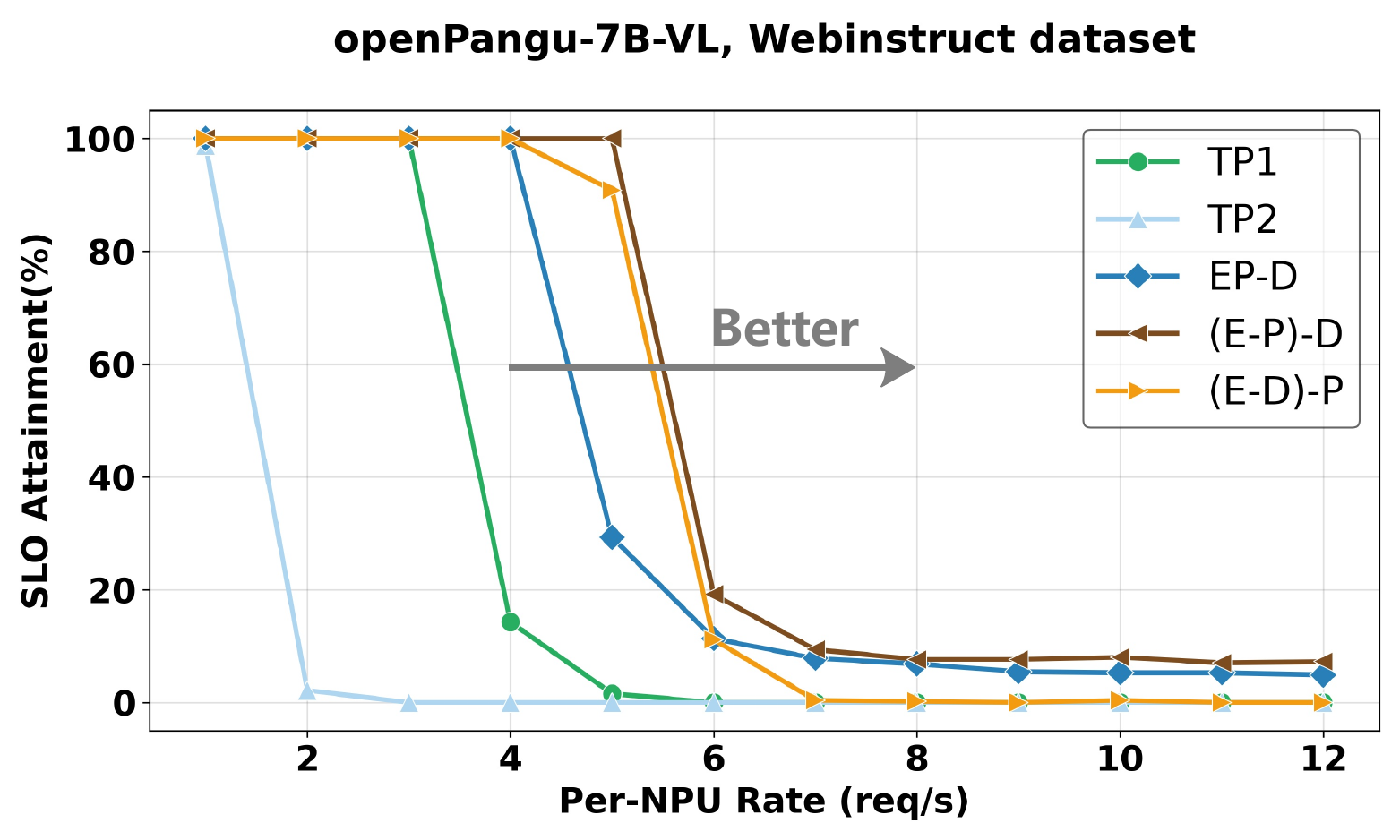}
    \hspace{0.1in}
    \includegraphics[width=0.4\textwidth]{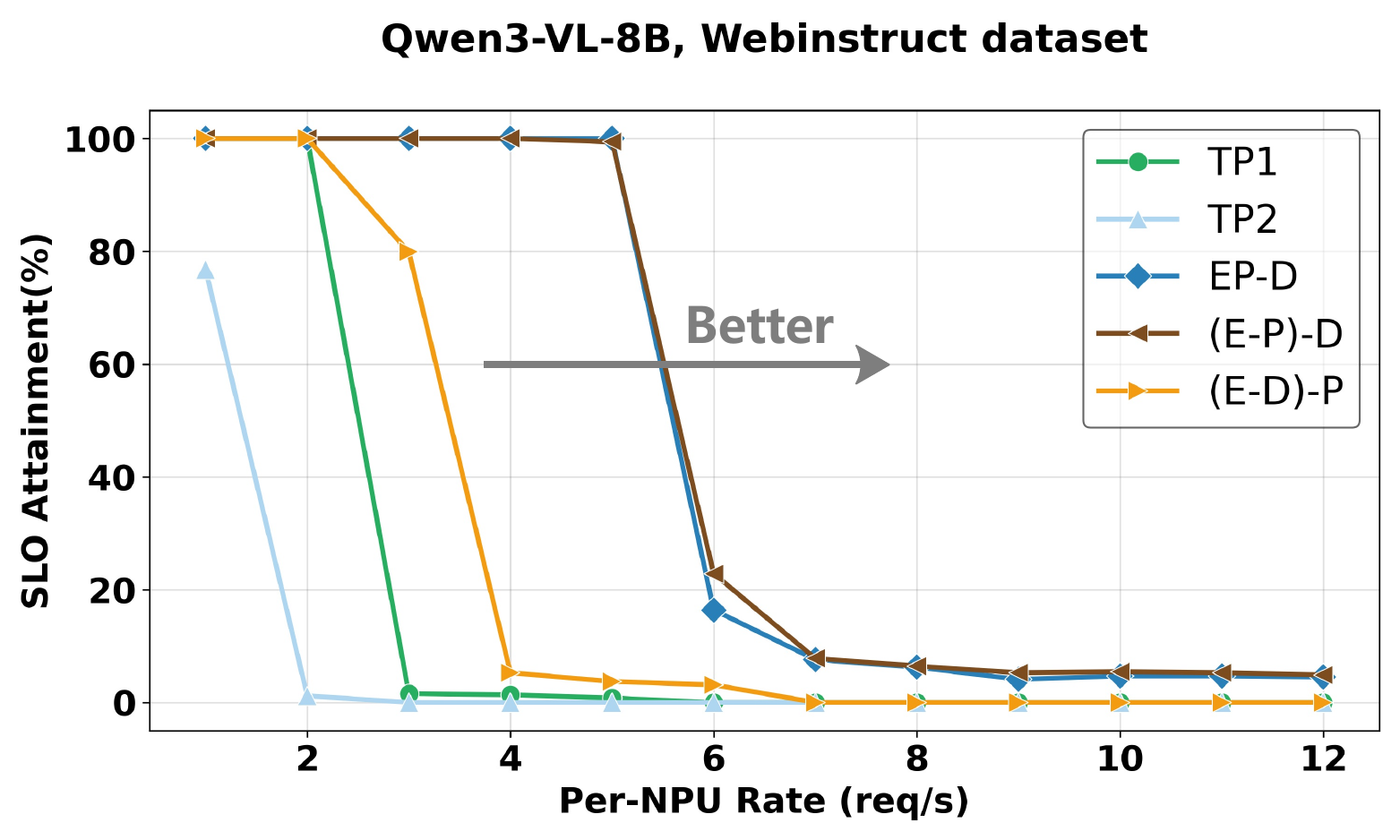}
    }
    \par\noindent\setlength{\parskip}{0pt}  
    \vspace{-5pt}  
    \subfigure{
    \includegraphics[width=0.4\textwidth]{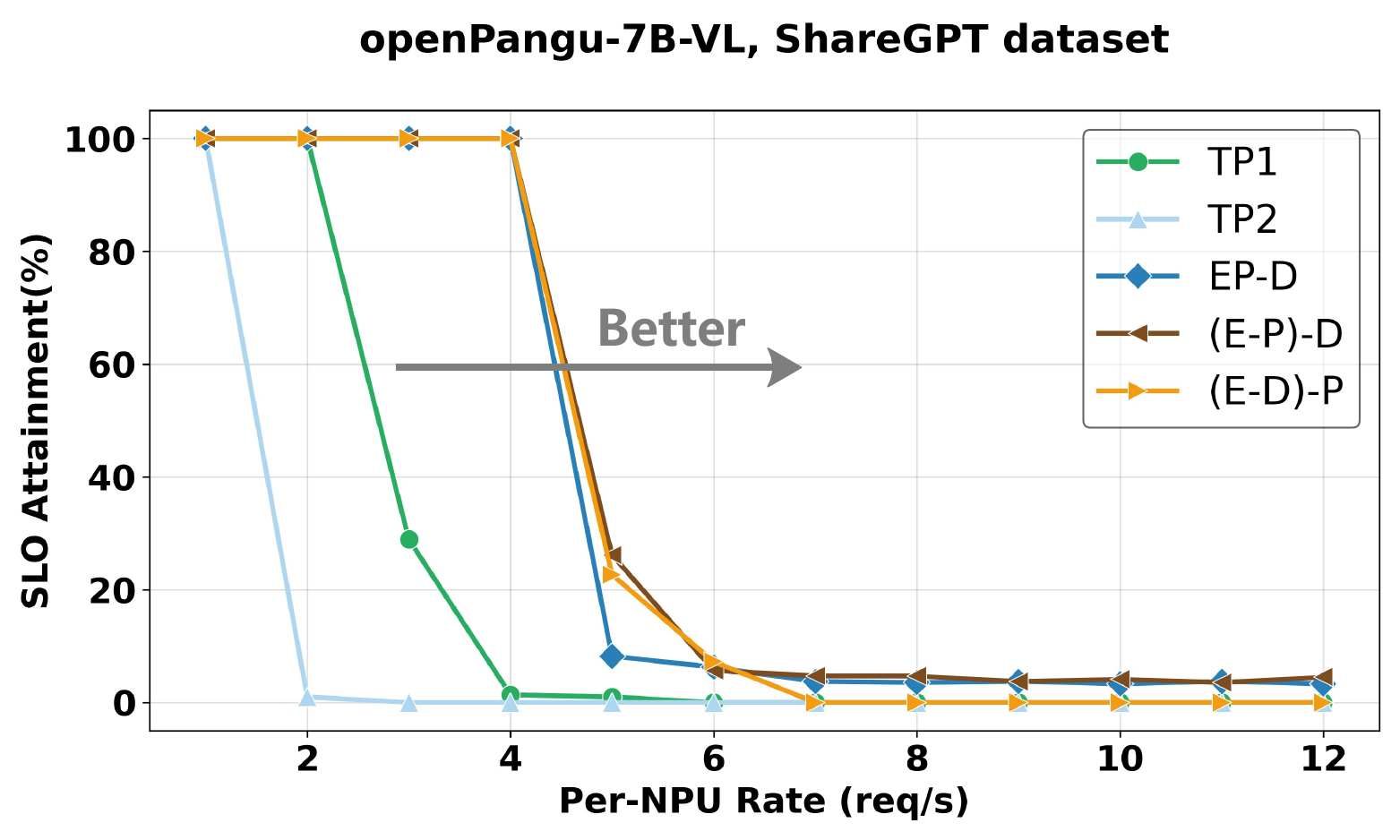}
    \hspace{0.1in}
    \includegraphics[width=0.4\textwidth]{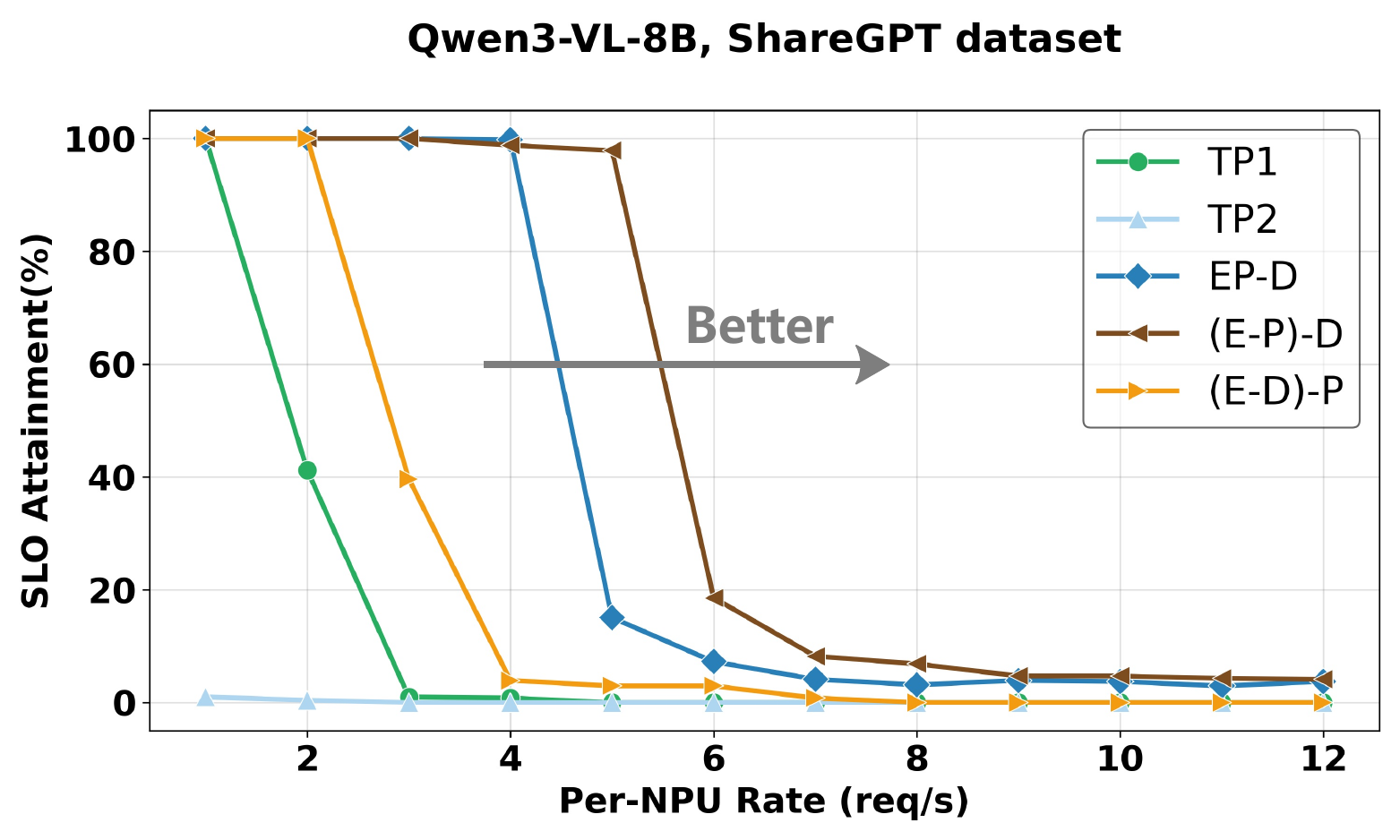}
    }
    \caption{Comparison of SLO attainment rate between Decode-stage disaggregated and monolithic deployments. 
    The (E-P)-D deployment, which further disaggregates and co-locates the E and P based on Decode-stage disaggregation, 
    consistently achieves a higher SLO attainment rate than both the TP1 baseline and EP-D deployment.}
    \label{img:4_3:a}
\end{figure}

\begin{figure}[!h]
    \centering
    \subfigure{
    \includegraphics[width=0.4\textwidth]{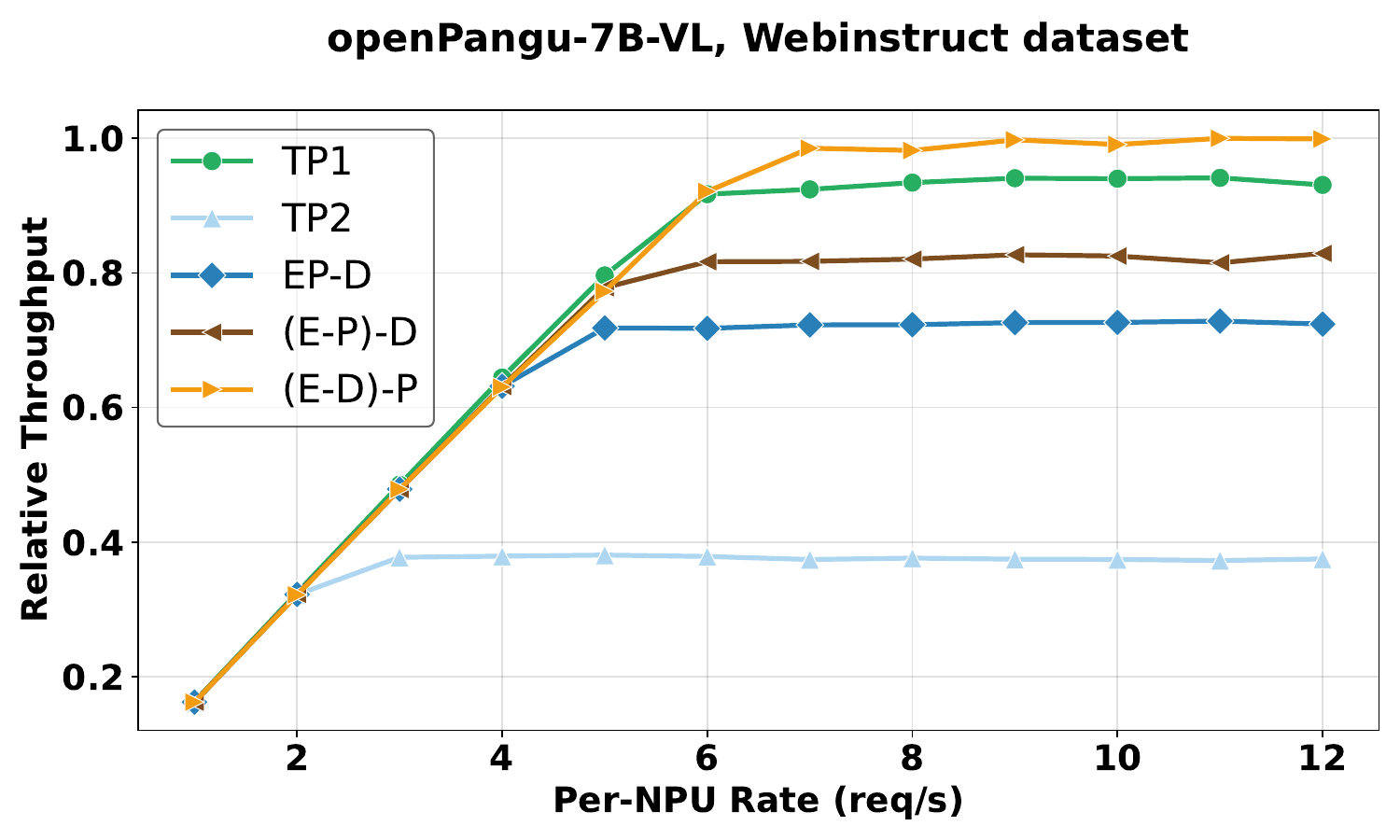}
    \hspace{0.1in}
    \includegraphics[width=0.4\textwidth]{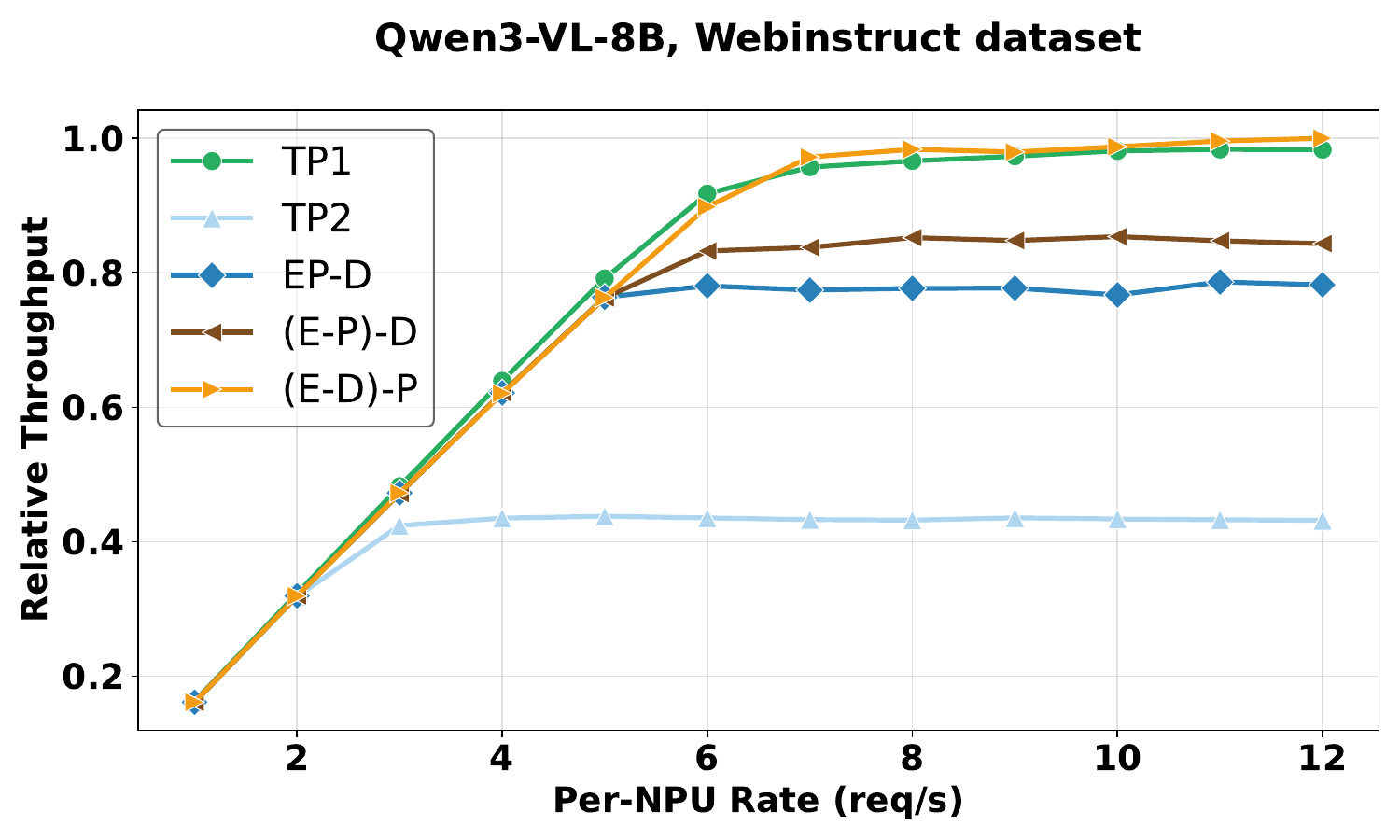}
    }
    \par\noindent\setlength{\parskip}{0pt}  
    \vspace{-5pt}  
    \subfigure{
    \includegraphics[width=0.4\textwidth]{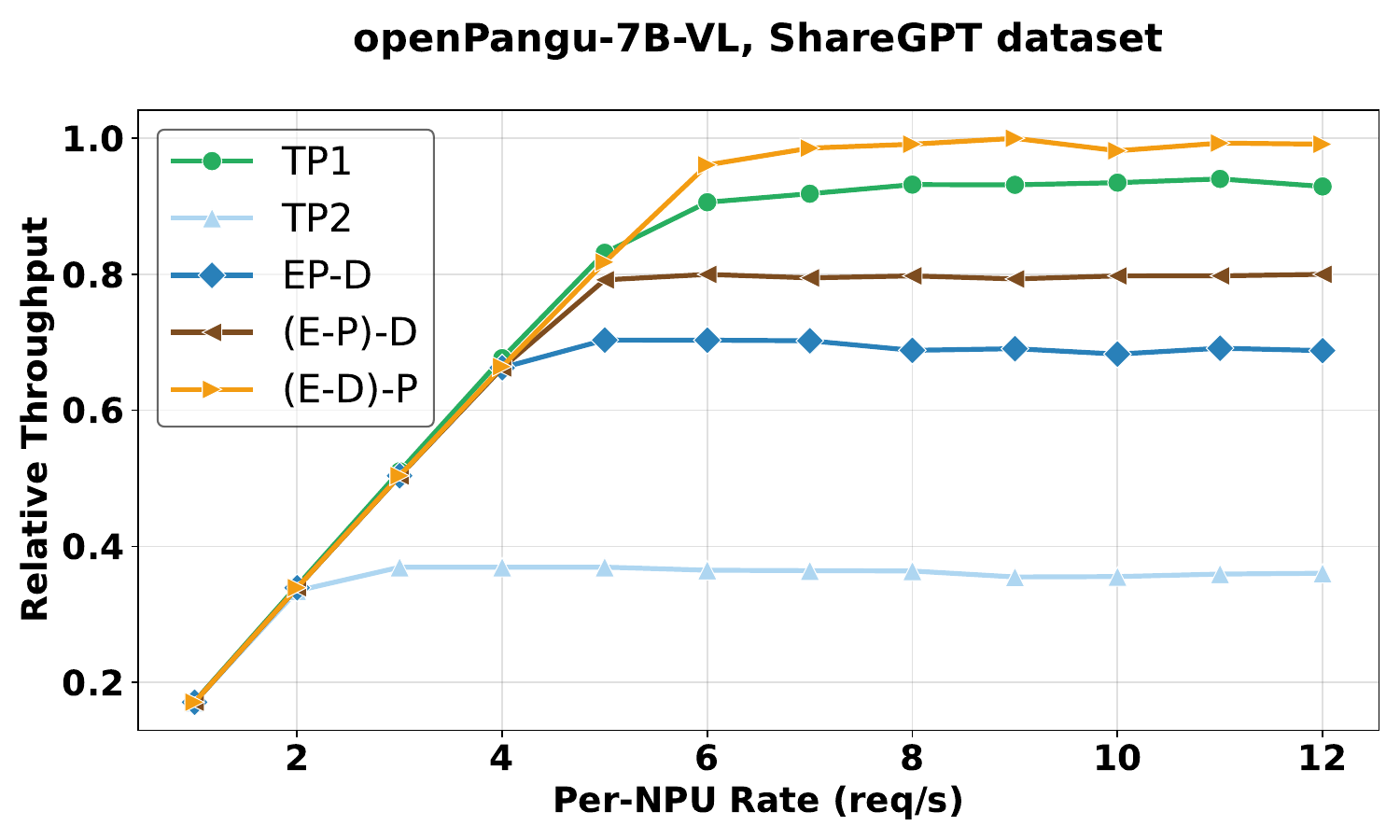}
    \hspace{0.1in}
    \includegraphics[width=0.4\textwidth]{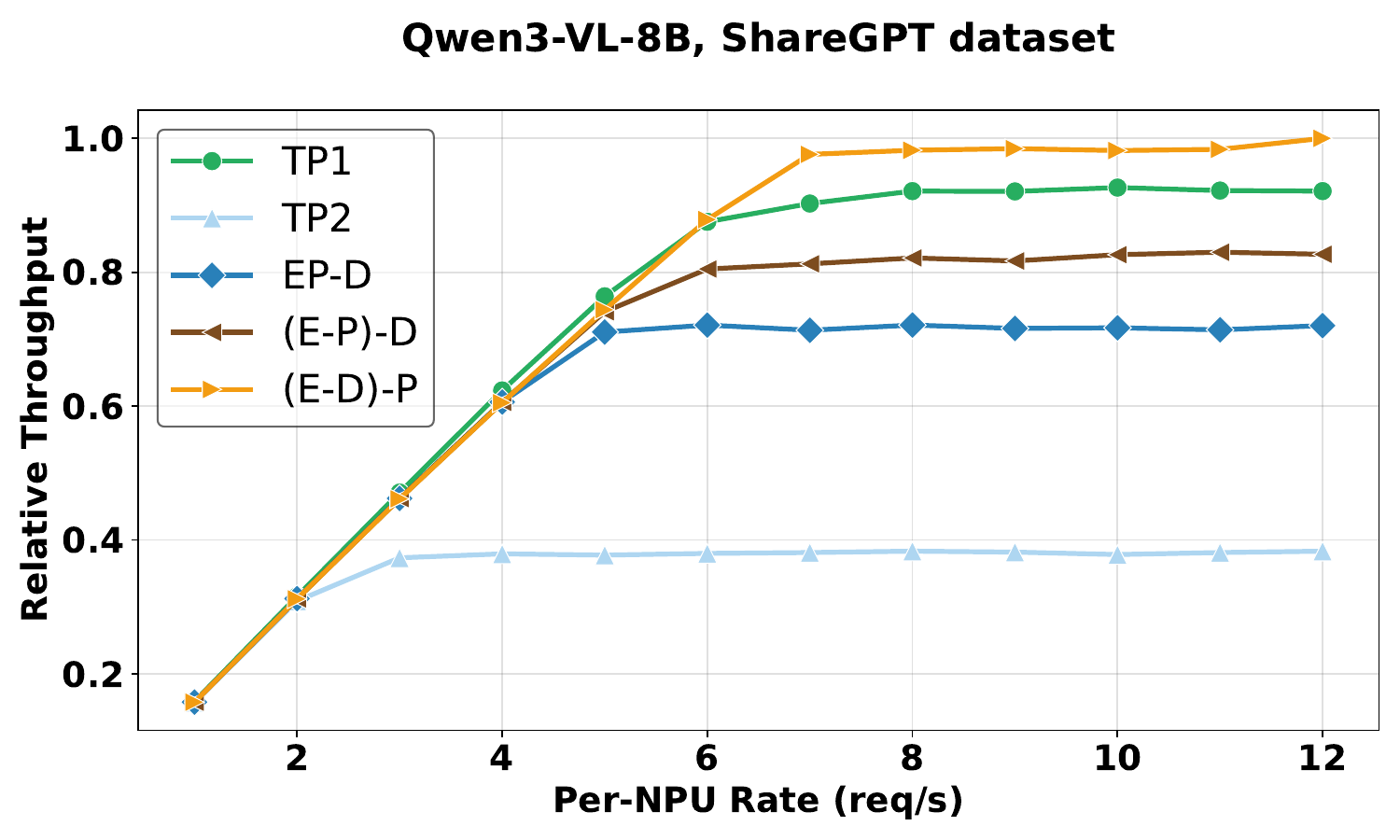}
    }
    \caption{Comparison of throughput performance between Decode-stage disaggregated and monolithic deployments. 
    The (E-D)-P deployment, which disaggregates and co-locates the E and D based on Decode-stage disaggregation, 
    outperforms the TP1 baseline and EP-D deployment in throughput.}
    \label{img:4_3:b}
\end{figure}

\begin{figure}[!h]
    \centering
    \subfigure{
    \includegraphics[width=0.4\textwidth]{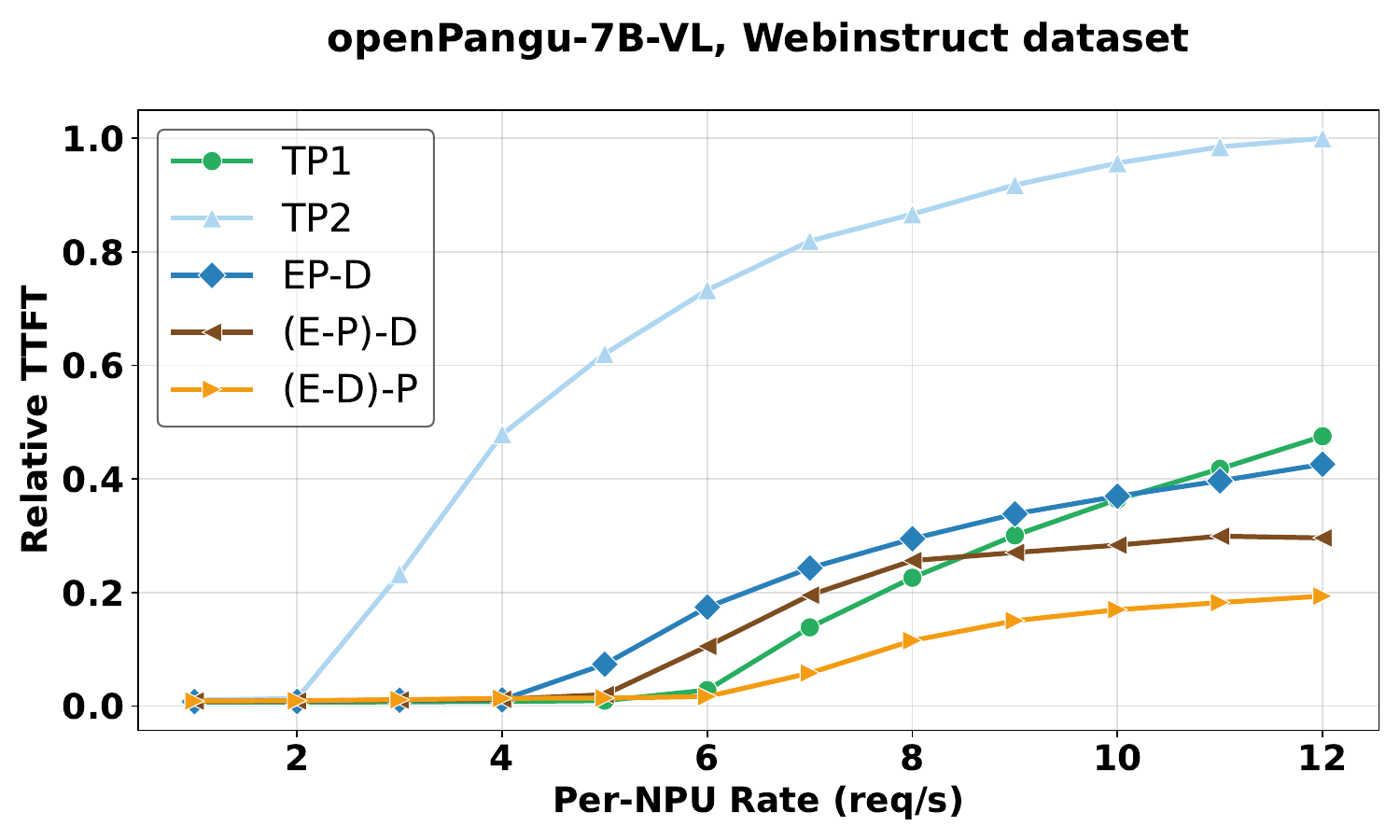}
    \hspace{0.1in}
    \includegraphics[width=0.4\textwidth]{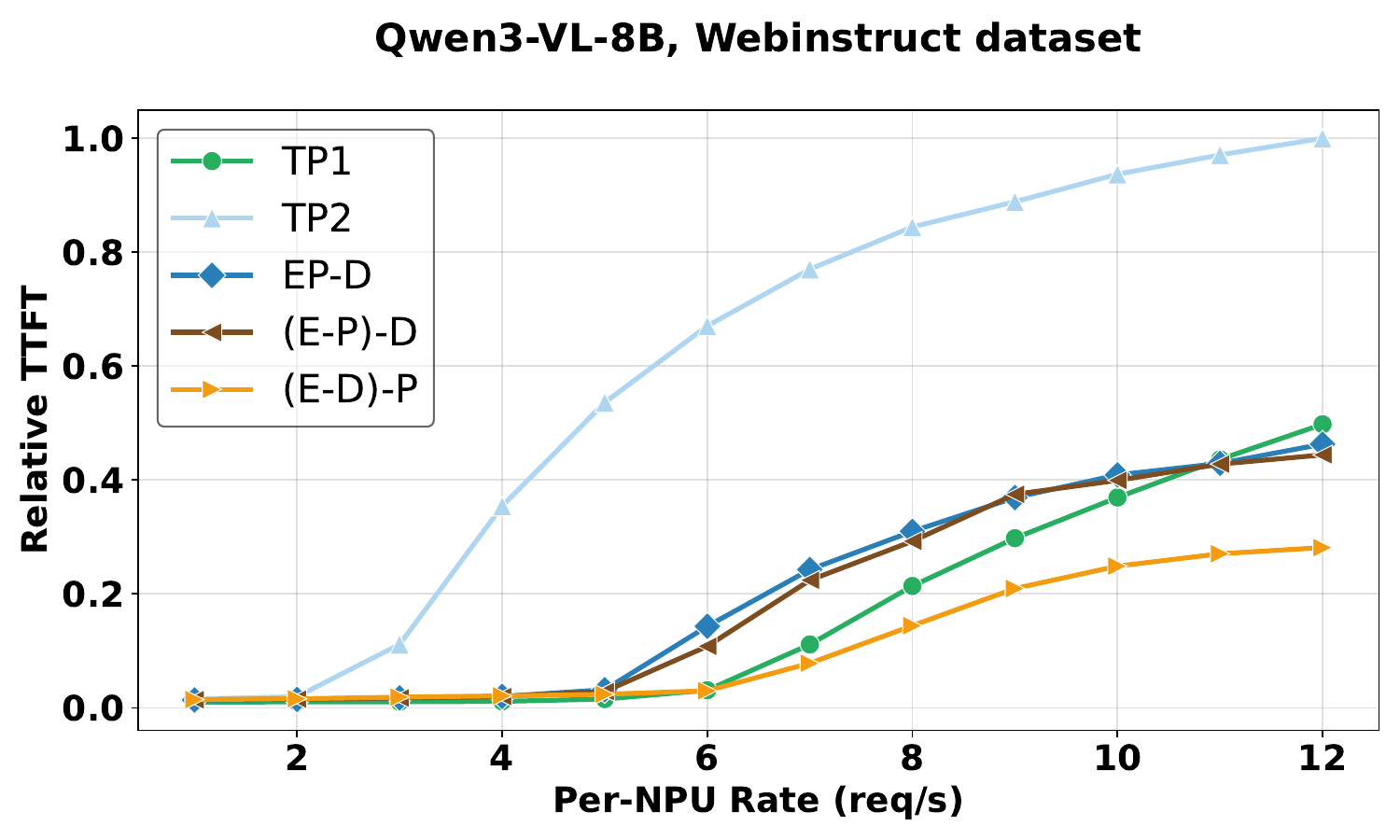}
    }
    \par\noindent\setlength{\parskip}{0pt}  
    \vspace{-5pt}  
    \subfigure{
    \includegraphics[width=0.4\textwidth]{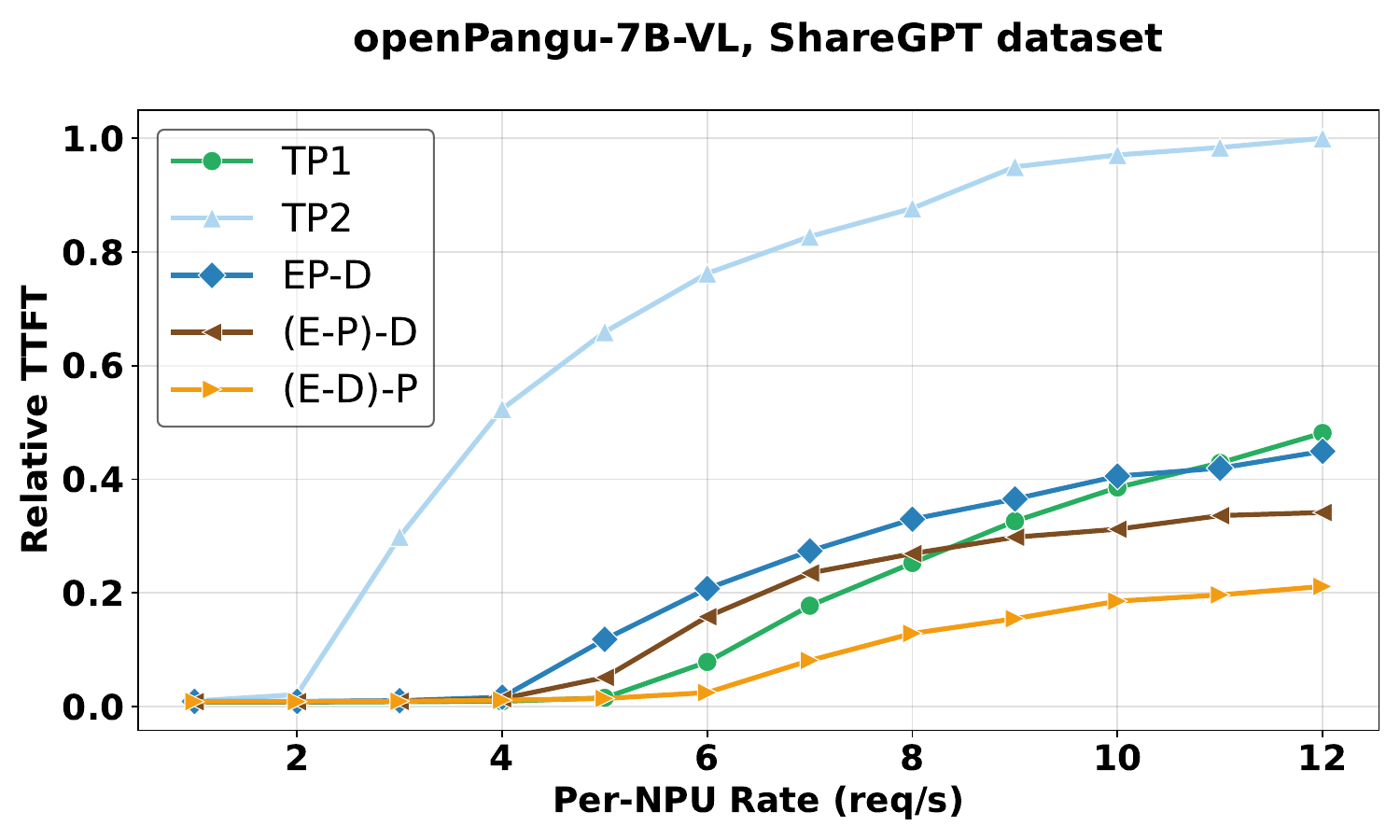}
    \hspace{0.1in}
    \includegraphics[width=0.4\textwidth]{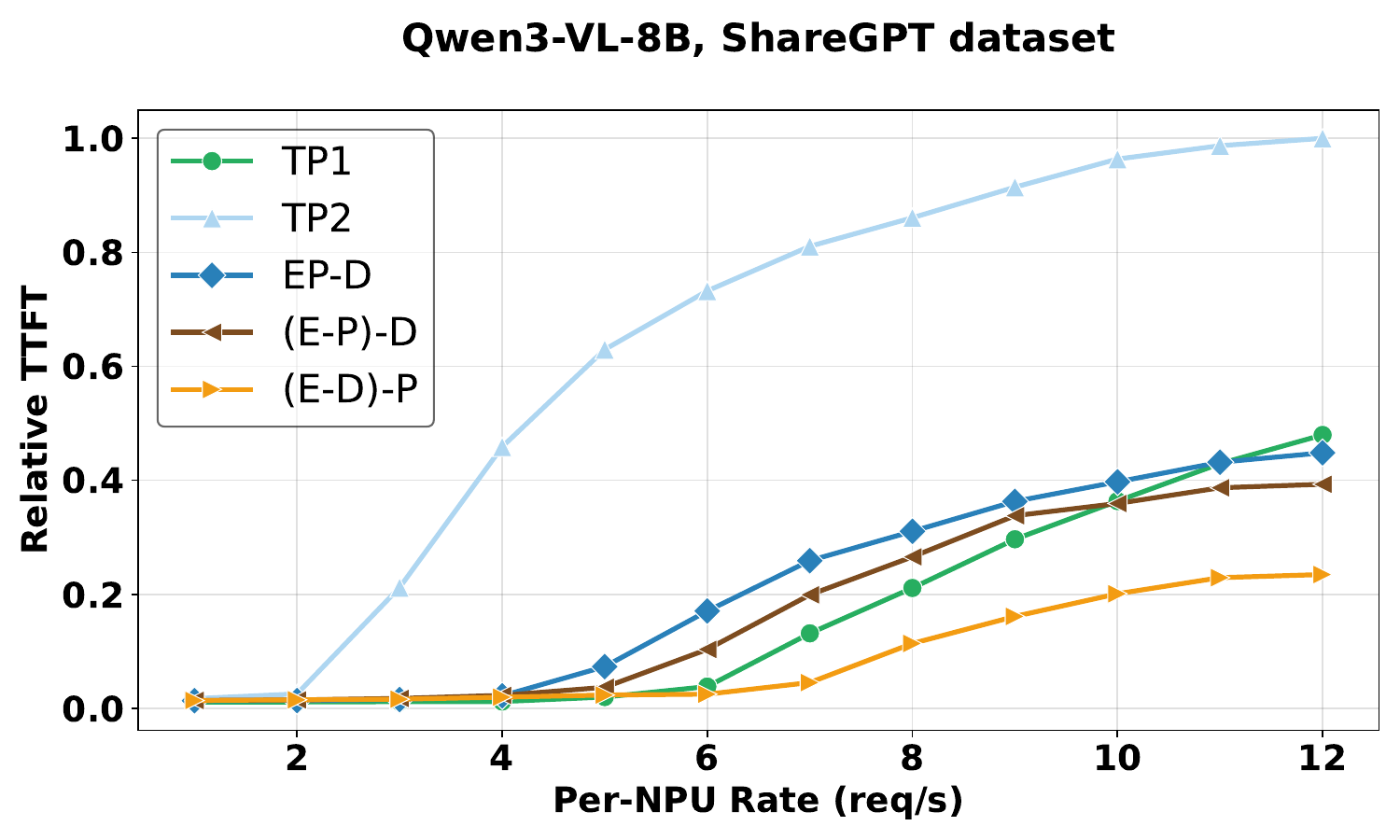}
    }
    \caption{Comparison of TTFT performance between Decode-stage disaggregated and monolithic deployments. 
    The (E-D)-P deployment, which disaggregates and co-locates the E and D based on Decode-stage disaggregation, 
    delivers lower TTFT latency than the TP1 baseline and EP-D deployment.}
   \label{img:4_3:c}
\end{figure}

\begin{figure}[!h]
    \centering
    \subfigure{
    \includegraphics[width=0.4\textwidth]{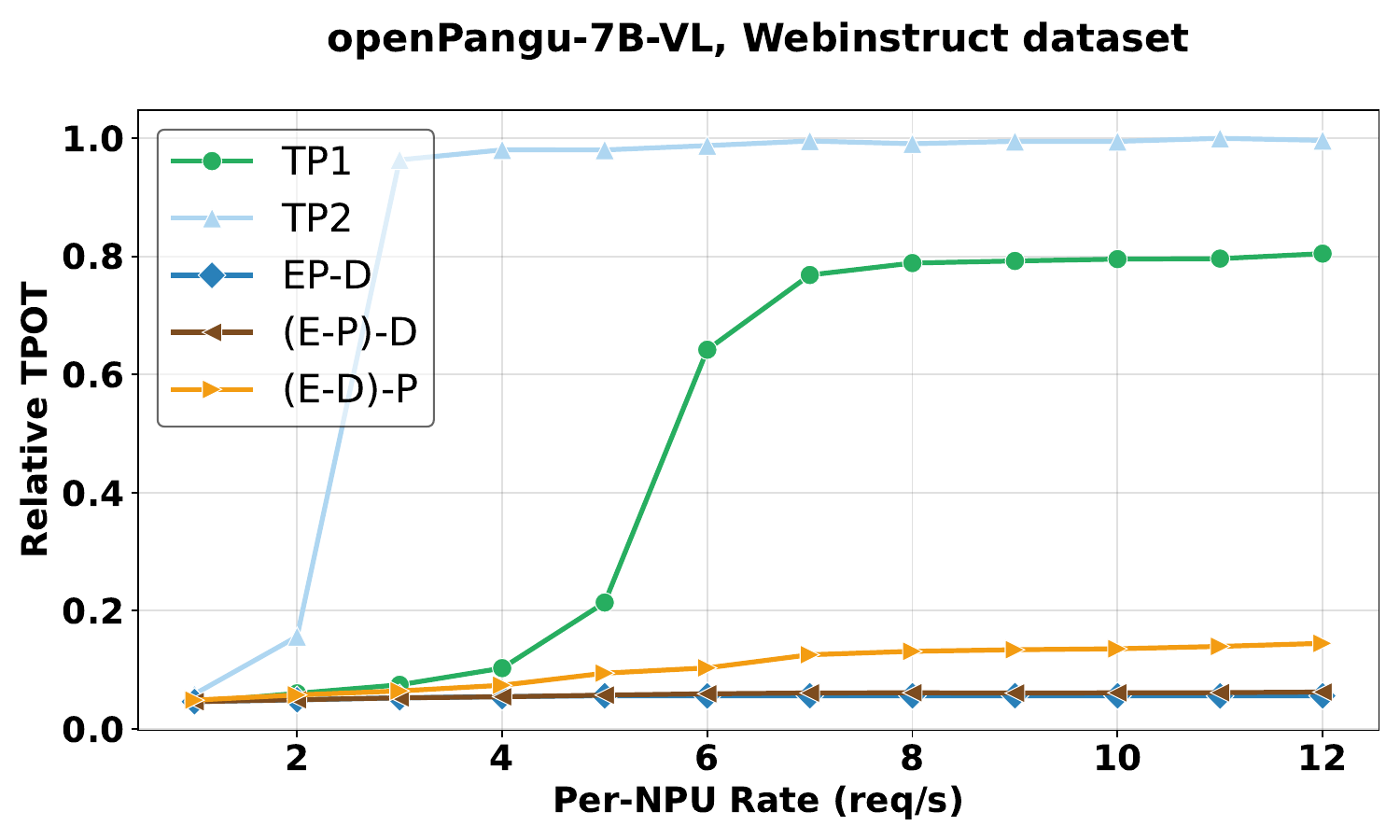}
    \hspace{0.1in}
    \includegraphics[width=0.4\textwidth]{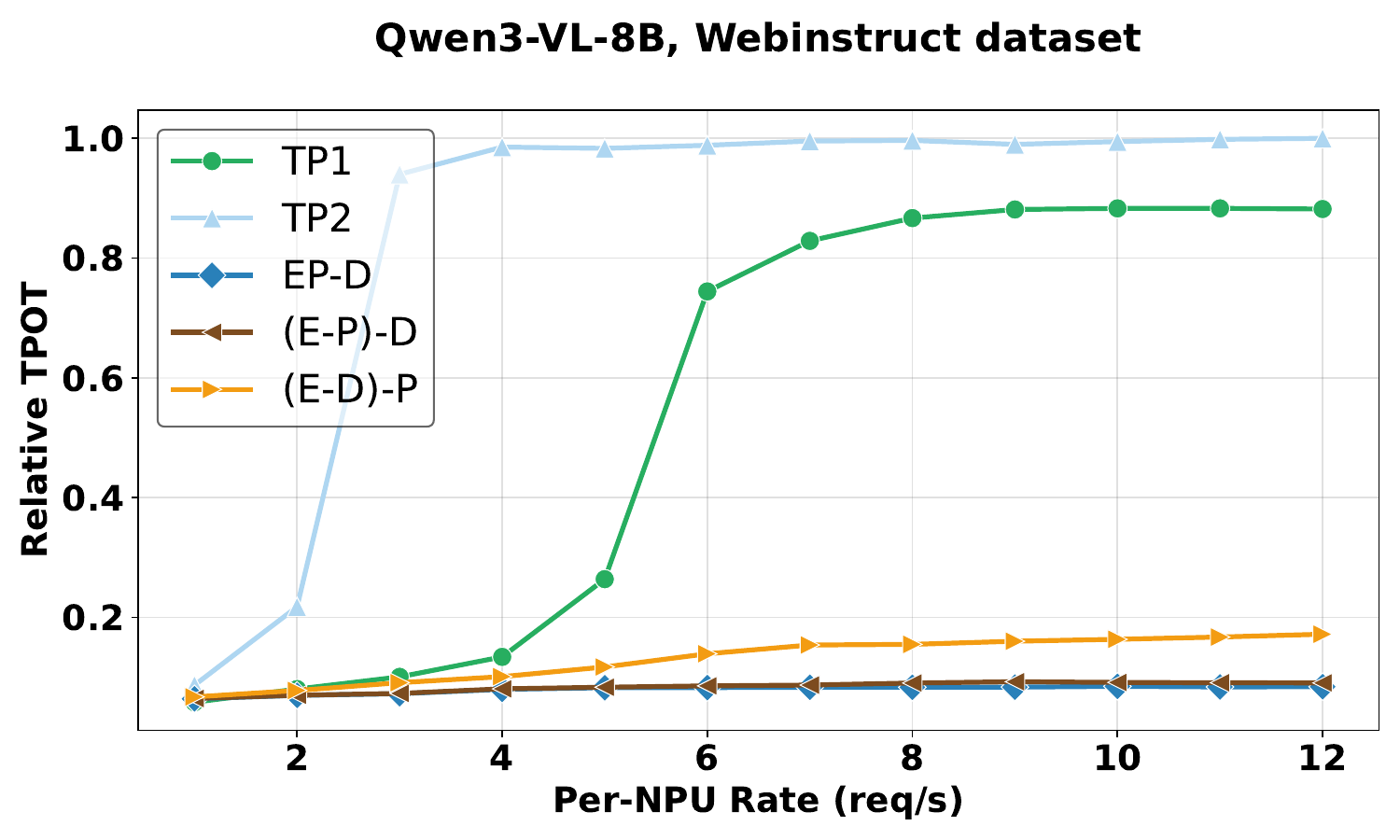}
    }
    \par\noindent\setlength{\parskip}{0pt}  
    \vspace{-5pt}  
    \subfigure{
    \includegraphics[width=0.4\textwidth]{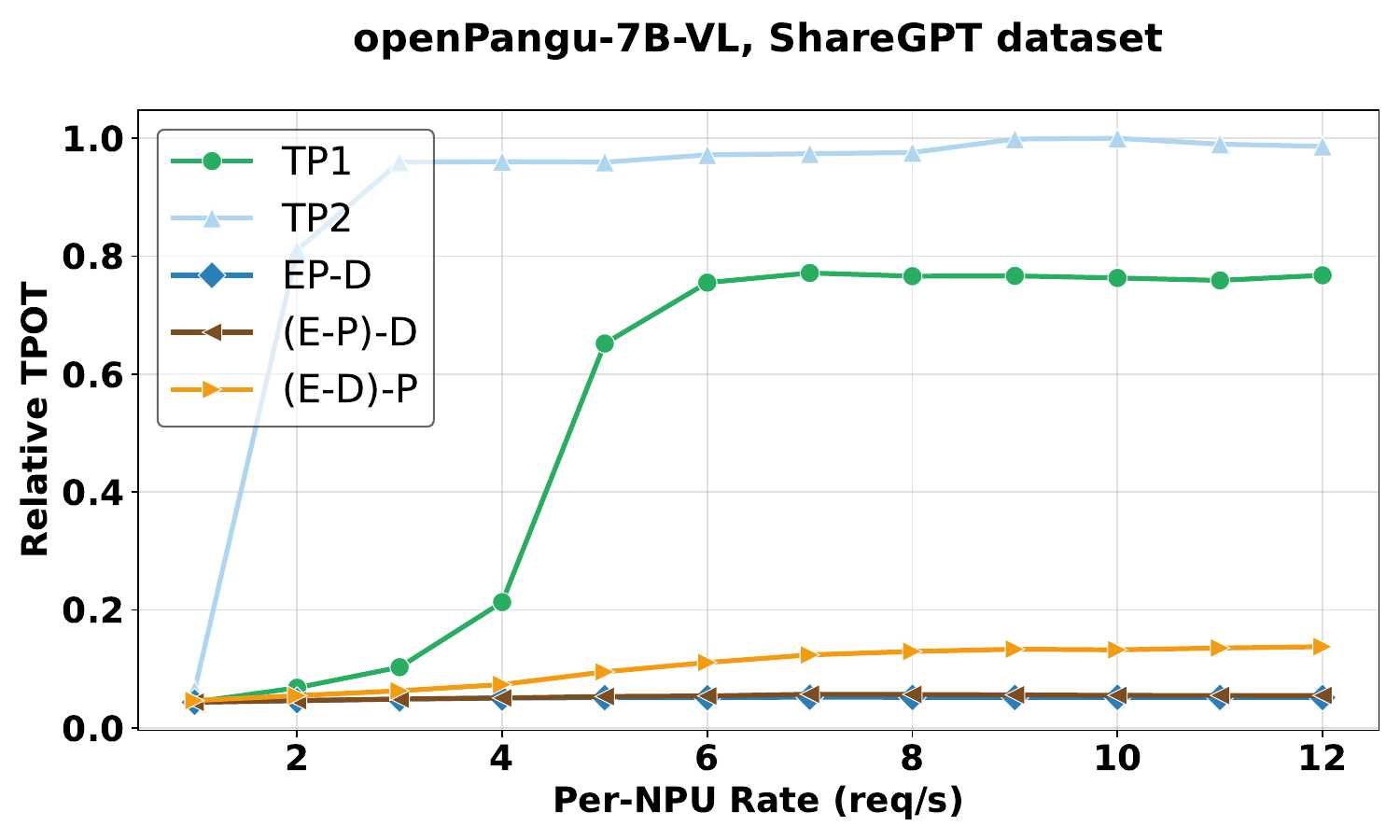}
    \hspace{0.1in}
    \includegraphics[width=0.4\textwidth]{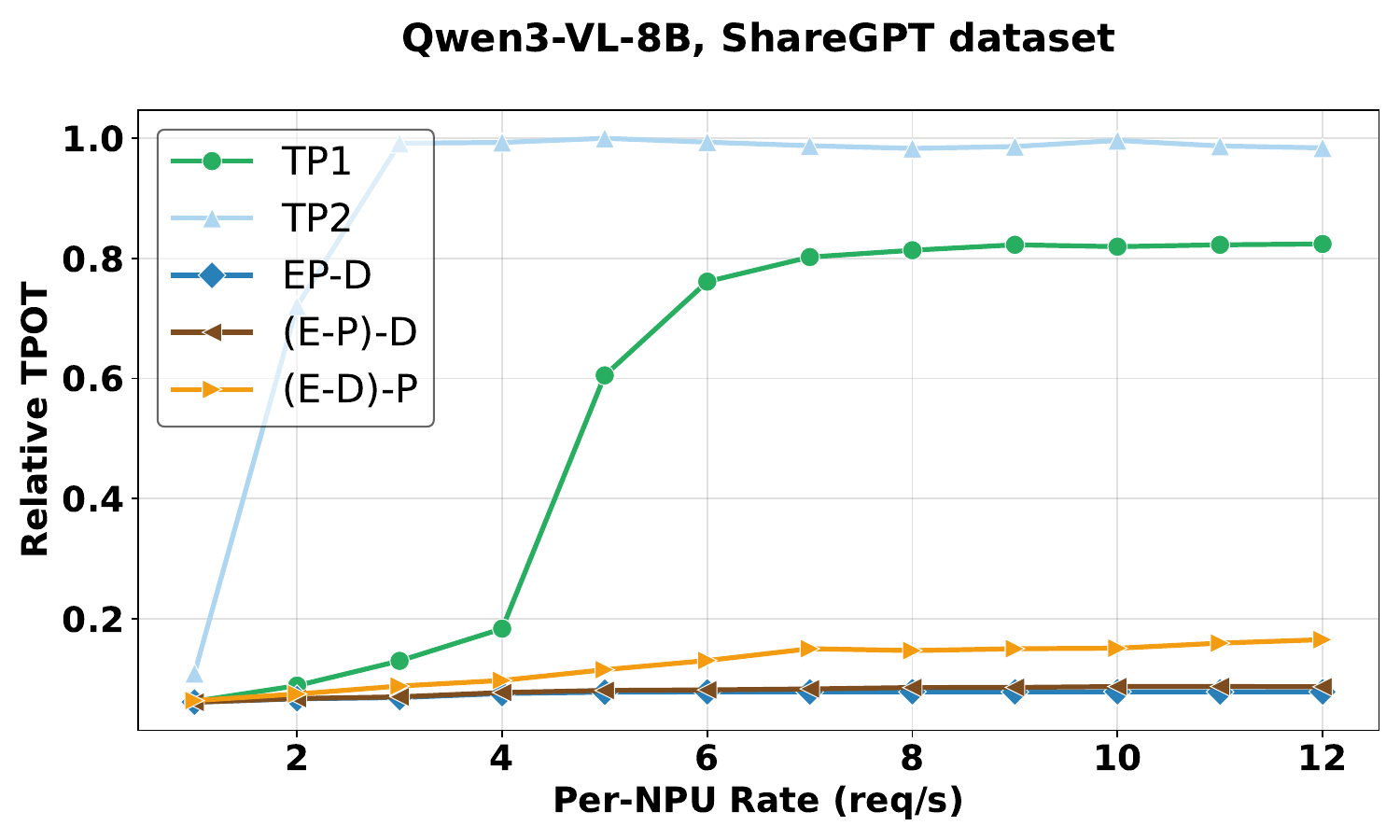}
    }
    \caption{Comparison of TPOT performance between Decode-stage disaggregated and monolithic deployments. 
    The (E-P)-D and EP-D deployments mitigate interference from the E and P stages by decoupling the D stage, 
    resulting in lower TPOT latency compared to other deployments.}
   \label{img:4_3:d}
\end{figure}

After examining the performance impact of Encode stage disaggregation, we now analyze the benefits of disaggregating the Decode stage.
\cref{img:4_3:a,img:4_3:b,img:4_3:c,img:4_3:d} illustrate the evolution of four key metrics, SLO attainment rate, throughput, TTFT, and TPOT, as the request injection rate increases, on both the VisualWebInstruct and ShareGPT-4o datasets. 
The comparison includes representative deployments such as TP1, TP2, EP-D, (E-P)-D, and (E-D)-P.

The experimental results show that all Decode-disaggregated deployments, such as EP-D, (E-P)-D, and (E-D)-P, exhibit significant TPOT advantages across both datasets and models. 
Under high concurrency at req/s=12, TPOT is reduced by 79.99-93.31\% relative to the TP1 baseline. 
This benefit arises because an independently deployed Decode stage no longer competes with Encode or Prefill for resources, thereby minimizing tail latency and demonstrating the central role of Decode disaggregation in stabilizing TPOT.

For TTFT, further decoupling Encode from Prefill, based on the Decode-disaggregated deployment, provides additional acceleration for first-token latency.
When Encode is decoupled from Prefill and co-located with Decode in the (E-D)-P deployment, TTFT decreases by 39.22-54.56\% compared with EP-D under high load. 
This improvement stems from the resource complementarity formed by the compute-intensive nature of Encode and the memory-intensive nature of Decode when co-located, thereby improving the execution efficiency of the Encode stage.
In contrast, (E-P)-D co-locates two compute-intensive stages, yielding slightly higher TTFT than (E-D)-P, though still significantly better than the non-co-located EP-D. This confirms that physical co-location of Encode and Prefill improves TTFT through spatial multiplexing.

Regarding TPOT, both (E-P)-D and EP-D deliver optimal performance due to the independently deployed Decode stage, consistently maintaining low generation latency. Although (E-D)-P also deploys Decode independently, its co-location with the Encode stage introduces minor resource contention during generation, causing slight TPOT degradation. However, this overhead remains small and is offset by the substantial TTFT gains.

Under SLO constraints requiring $TTFT \leq 2000ms$ and $TPOT \leq 50ms$, only (E-P)-D and EP-D satisfy both latency requirements when using openPangu-7B-VL. 
Among them, (E-P)-D not only matches EP-D in TPOT but also achieves higher effective throughput, improving by 57.37-69.48\% relative to EP-D, highlighting superior resource utilization and throughput scalability under high concurrency.
Under more strict SLO constraints of $TTFT < 800ms$ and $TPOT < 30ms$, experiments are conducted on the ShareGPT-4o dataset with the average per-card request rate fixed at 4 req/s. 
In this setting, EP-D achieves an SLO attainment rate of 59.57\% with an effective throughput of 294.68 tokens/s, whereas (E-P)-D improves the SLO attainment rate to 84.96\% and increases effective throughput to 420.16 tokens/s. 
Compared with EP-D, (E-P)-D still delivers a 42.58\% improvement in effective throughput under these stricter latency constraints.

In summary, independent deployment of the Decode stage is essential for achieving stable, low-tail TPOT. Building on this foundation, further decoupling the Encode stage and choosing appropriate co-location strategies with Decode or Prefill enables additional TTFT or throughput gains, providing flexible performance tuning across workload levels and SLO.

\subsection{Benefits of Full Encode-Prefill-Decode Disaggregation}
After separately analyzing the effects of Encode and Decode disaggregation, we now examine the synergistic behavior and performance benefits of fully disaggregating all three stages including Encode, Prefill, and Decode. Experiments are conducted on the ShareGPT-4o dataset with a fixed request rate of 10 req/s to compare TTFT, TPOT, and SLO attainment of different deployment strategies for the openPangu-7B-VL model. The results are summarized in Table \ref{tab:4_4}.

\begin{table}[t]
    \centering
    \footnotesize
    \setlength{\tabcolsep}{4.5pt}
    \caption{Performance comparison of different deployments for openPangu-7B-VL under high-load conditions, 10 req/s.}
    \small
    \begin{tabular}{cccccc}
    \toprule
Deployment & NPUs Number & TTFT(ms) & TPOT(ms) & SLO Attainment Rate & Per-NPU Effective Throughput \\\midrule
TP1$\times$2      & 2            & 658.27                 & 95.56                & 2.15\%              & 13.38               \\
(E-PD)$\times$2   & 2            & 548.32                 & 62.22                & 3.13\%              & 19.70               \\
EP-D              & 2            & 5523.82                & 27.31                & 8.20\%              & 21.54               \\
(E-P)-D           & 2            & 2386.85                & 28.40                & 26.17\%             & 77.36               \\
(E-D)-P           & 2            & 651.86                 & 50.71                & 22.66\%             & 69.18               \\
E-P-D             & 3            & 557.89                 & 28.92                & 94.34\%             & 192.70              \\\bottomrule    
\end{tabular}
    \label{tab:4_4}
\end{table}

Under high-load conditions at 10 req/s and stringent SLO constraints requiring $TTFT \leq 2000ms$ and $TPOT \leq 50ms$, 
only four deployments, including EP-D, (E-P)-D, (E-D)-P, and E-P-D, can meet the SLO for a portion of requests. 
Among them, E-P-D delivers the best performance, achieving a 94.34\% SLO attainment rate, the highest of all deployments.
Its per-NPU effective throughput is 7.95 times that of EP-D, indicating that fully disaggregating all three stages enables substantially higher request-serving capacity under strict SLO constraints. These results demonstrate that the E-P-D deployment is particularly effective at improving the fraction of requests that meet tight latency constraints.

\subsection{Comprehensive Analysis of EPD-Disaggregated Deployments}
\begin{figure}[htbp]
  \centering
  \subfigure[TTFT distribution scatter plot of all requests.
Higher-performing deployments cluster in the low-TTFT region with more successful requests.
Under high load conditions, (E-P)-D, (E-D)-P, and EP-D concentrate in the low-TTFT region, outperforming other deployments.
]{
  \label{img:4_4:a}
  \includegraphics[width=0.75\linewidth]{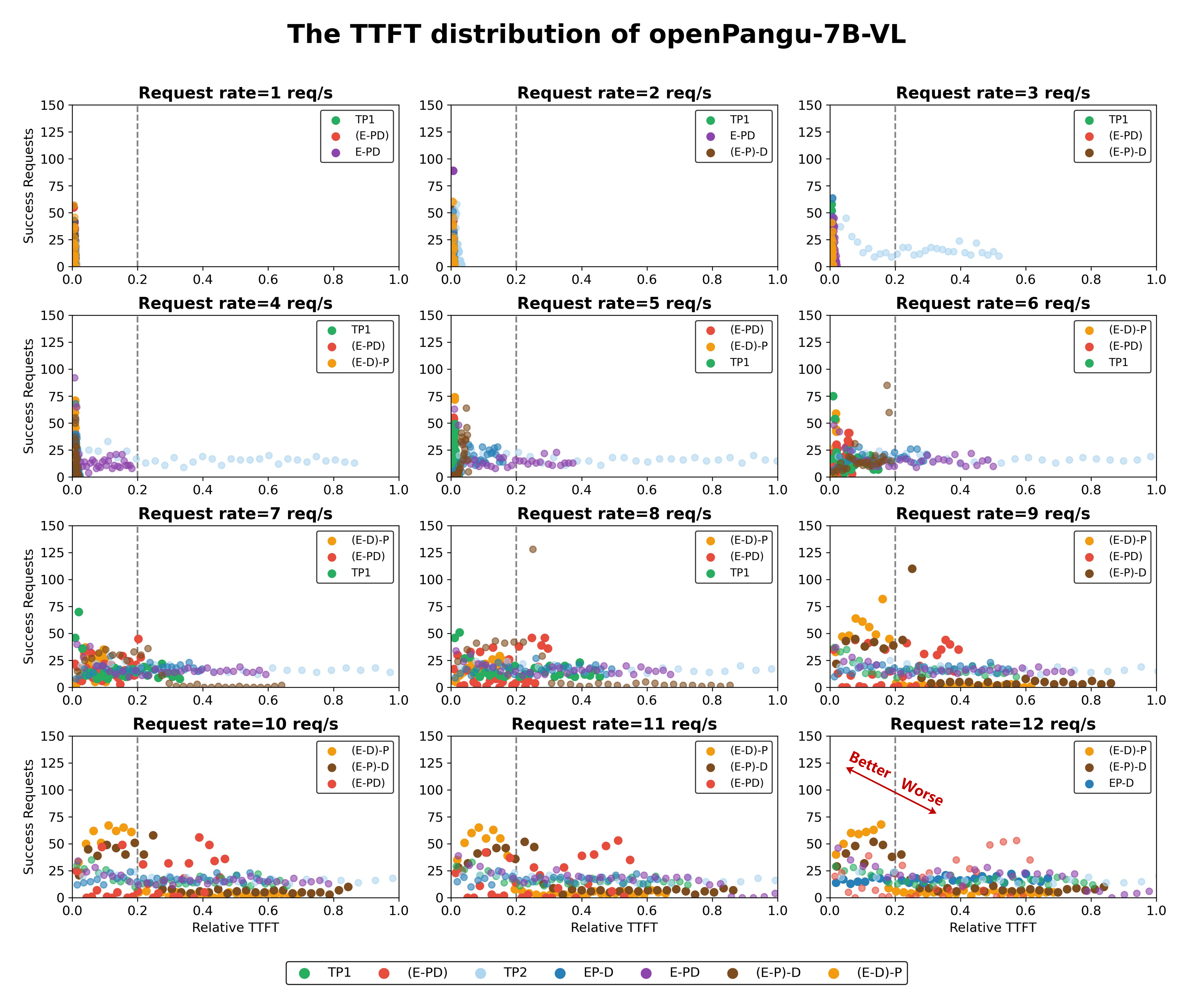}}
  \\
  \subfigure[TPOT distribution scatter plot of all requests.
Higher-performing deployments cluster in the low-TPOT region with more successful requests.
Under high load conditions, EP-D, (E-P)-D, and (E-D)-P concentrate in the low-TPOT region, outperforming other deployments.
]{
  \label{img:4_4:b}
  \includegraphics[width=0.75\linewidth]{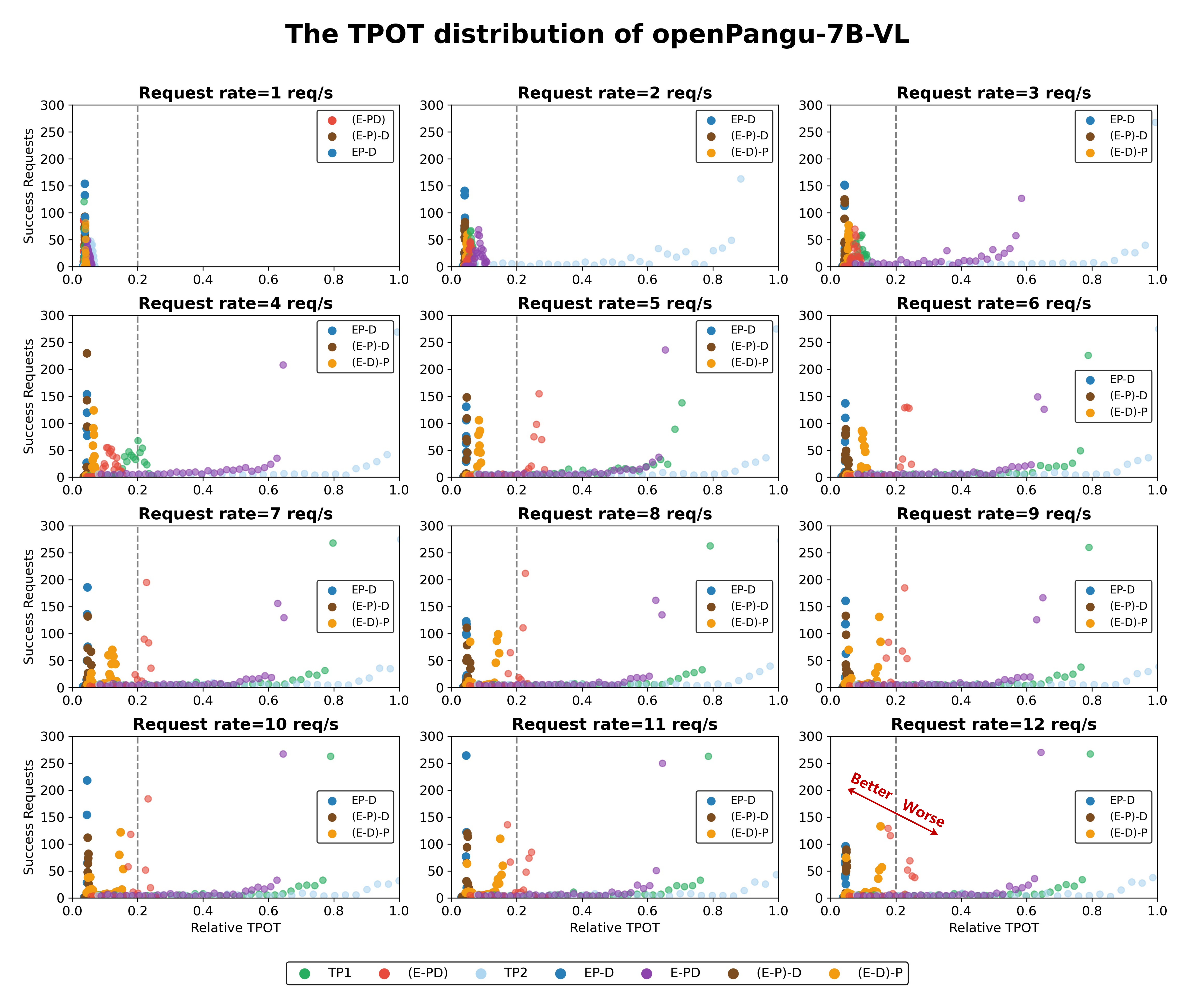}}
  \caption{TTFT and TPOT scatter plots of openPangu-7B-VL across deployments under increasing request rates.
The subfigure legend highlights the top three deployments at each request rate.}
  \label{img:4_4}
\end{figure}

To further examine system behavior after disaggregating the Encode and Decode stages, Figure \ref{img:4_4} presents scatter plots of TTFT and TPOT distributions for the openPangu-7B-VL model under different request rates on the ShareGPT-4o dataset. These visualizations reveal how performance evolves across deployments under high concurrency from a fine-grained, request-level perspective.

For TTFT distribution, as the request injection rate increases, TP2 is the first to hit the system's processing limit, exhibiting clear queuing and rapidly rising TTFT. 
E-PD and TP1 subsequently enter overload. Under the high-load condition of 12 req/s, only (E-P)-D, (E-D)-P, and EP-D deployments maintain a relatively low TTFT range, while all others experience backlog and shift into the high-latency region. Notably, (E-D)-P achieves the best TTFT performance due to the complementary resource usage between its disaggregated Encode and Decode stages, confirming the effectiveness of Encode-Decode co-location for first-token latency optimization.

A similar pattern appears in the TPOT distribution. 
TP2 is the first to exhibit backlog, followed by E-PD, TP1, and (E-P)-D, indicating that PD monolithic deployments without Decode disaggregation are more susceptible to Decode-stage queuing and thus incur higher TPOT. 
At 12 req/s, all Decode-disaggregated deployments, like EP-D, (E-P)-D, and (E-D)-P, maintain low TPOT and clearly outperform monolithic deployments. 
Because (E-D)-P co-locates Encode and Decode, it experiences minor resource contention during generation, resulting in slightly higher TPOT than EP-D and (E-P)-D, though performance remains stable. 
Overall, under high concurrency, Decode-disaggregated designs consistently deliver the lowest output latency, demonstrating that decoupling Prefill and Decode effectively isolates the generation stage from new-request interference and ensures stable TPOT.

\subsection{Beneficial Scenarios for the EPD-Disaggregated Deployments}

\begin{figure}[htbp]
    \begin{center}
\centerline{\includegraphics[width=1.0\columnwidth]{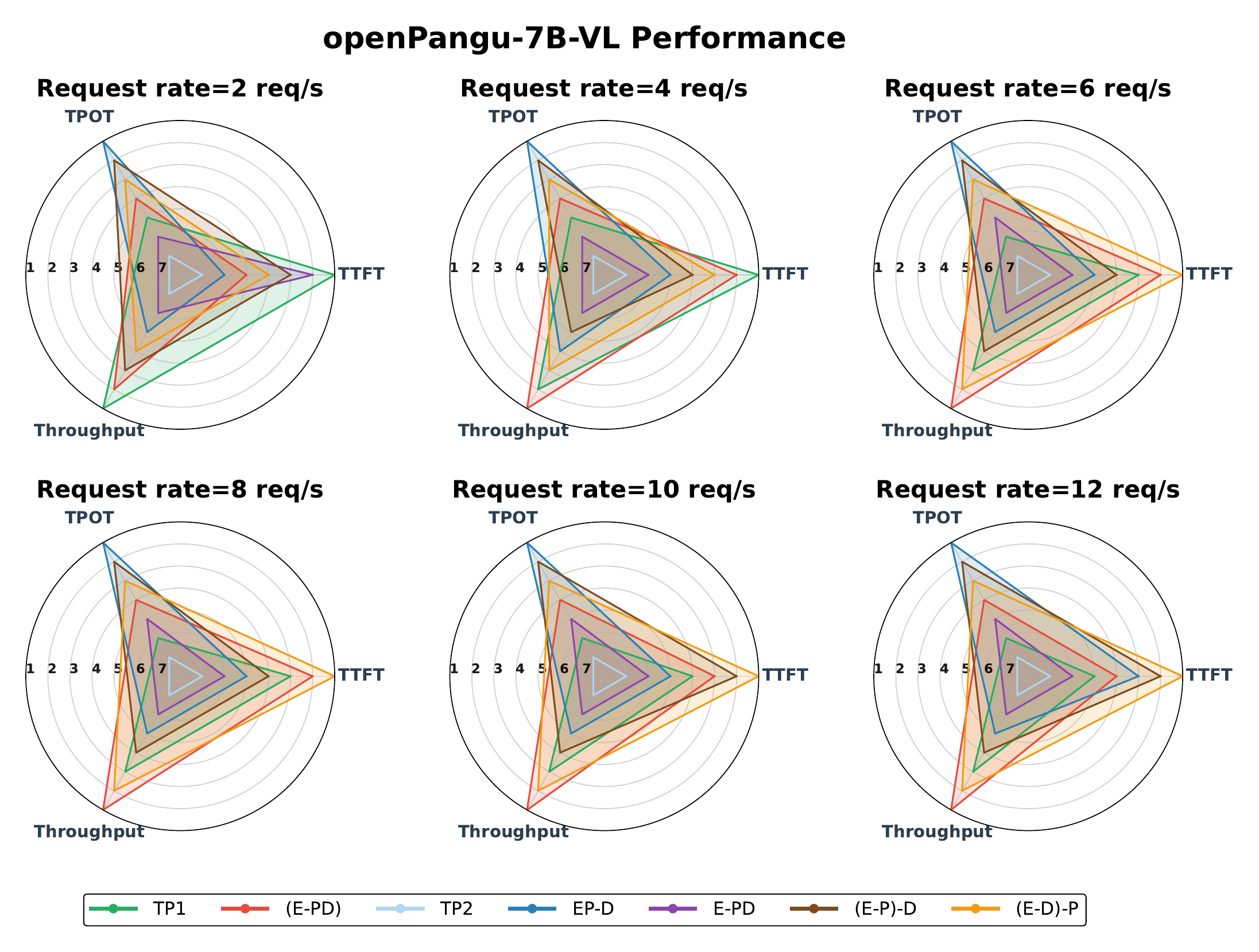}}
    \caption{TTFT, TPOT, and throughput performance of openPangu-7B-VL across deployments under varying request rates.
This figure presents a radar chart where the 1-7 concentric circles indicate the ranking of deployments based on performance, with 1 representing the best.
Under high-load scenarios, the EP-D, (E-D)-P, and (E-PD) deployments perform best on the TPOT, TTFT, and throughput, respectively.}
    \label{img:4_5}
    \end{center}
\end{figure}

To clarify the applicability and performance advantage regions of different EPD-disaggregated deployments, Figure \ref{img:4_5} presents radar charts of TTFT, TPOT, and throughput for the openPangu-7B-VL model under varying request rates on the ShareGPT-4o dataset. These charts intuitively reveal the performance gains and the advantage regions of each deployment across different levels of concurrency pressure.

As system load increases, the performance advantages and limitations of different deployments become increasingly distinct. Choosing an appropriate EPD-disaggregated strategy therefore requires aligning deployment decisions with specific SLO priorities:

\paragraph{High Performance: Low TTFT and Low TPOT.} 
For scenarios demanding both fast first-token response and stable generation latency, (E-P)-D offers the most balanced performance. As shown in Figure \ref{img:4_5}, it maintains low TTFT and TPOT even under high concurrency, reflecting the benefits of three-stage disaggregation combined with selective co-location. This deployment is well suited for latency-critical production workloads with strict SLO constraints.

\paragraph{Fast Response for First-token: Low TTFT with Moderate TPOT Tolerance.}
When minimizing first-token latency is the primary objective and moderate TPOT is acceptable, (E-D)-P is preferable. Independent deployment of Encode significantly accelerates TTFT, although co-location with Decode introduces minor contention during generation, yielding slightly higher TPOT compared with EP-D. This deployment fits applications where rapid initial response is essential, such as short-text generation tasks.

\paragraph{Maximizing Throughput: Loose TTFT/TPOT Constraints.}
For workloads prioritizing throughput over strict latency metrics, (E-PD) provides clear advantages. By decoupling Encode and co-locating it with the PD stage on the same hardware, this deployment achieves substantial throughput gains, as shown in Figure\ref{img:4_5}, despite being unable to meet tight SLO constraints for TTFT and TPOT. This makes it suitable for high-load, multi-user scenarios or RL post-training inference pipelines.

In summary, EPD-disaggregation strategies present complementary trade-offs among TTFT, TPOT, and throughput. (E-P)-D is ideal for meeting stringent latency SLOs, (E-D)-P excels when TTFT is the dominant requirement, and (E-PD) maximizes throughput under relaxed latency constraints. Such SLO-driven, fine-grained deployment selection enables \emph{EPD-Serve} to balance resource utilization and performance effectively across diverse workloads.

\section{Conclusion}
This paper proposes \emph{EPD-Serve},
which decouples the multimodal inference process into Encode, Prefill, and Decode stages.
By enabling staged scheduling, efficient cross-stage tensor transmission, and hardware-level co-location with spatial multiplexing,
\emph{EPD-Serve} overcomes the concurrency and resource-utilization bottlenecks inherent in monolithic inference architectures.
Experimental results on Ascend show that under high-concurrency multimodal workloads, \emph{EPD-Serve} substantially improves throughput while reducing TTFT and TPOT, all within strict SLO constraints. These findings validate the effectiveness of stage decoupling and compute-resource complementarity. Further analyses indicate that Encode-stage disaggregation and co-location primarily optimize TTFT, whereas Decode-stage disaggregation is critical for stabilizing TPOT, providing practical guidance for selecting deployment strategies under different SLO constraints.
Overall, \emph{EPD-Serve} demonstrates that combining logical isolation with physical co-location offers a scalable and efficient optimization path for multimodal inference systems.

\bibliographystyle{unsrt}
\bibliography{ref}

\begin{thebibliography}{10}

\bibitem{radford2021learning}
Alec Radford, Jong~Wook Kim, Chris Hallacy, Aditya Ramesh, Gabriel Goh, Sandhini Agarwal, Girish Sastry, Amanda Askell, Pamela Mishkin, Jack Clark, et~al.
\newblock Learning transferable visual models from natural language supervision.
\newblock In {\em International conference on machine learning}, pages 8748--8763. PmLR, 2021.

\bibitem{wu2023visual}
Chenfei Wu, Shengming Yin, Weizhen Qi, Xiaodong Wang, Zecheng Tang, and Nan Duan.
\newblock Visual chatgpt: Talking, drawing and editing with visual foundation models.
\newblock {\em arXiv preprint arXiv:2303.04671}, 2023.

\bibitem{yang2023mm}
Zhengyuan Yang, Linjie Li, Jianfeng Wang, Kevin Lin, Ehsan Azarnasab, Faisal Ahmed, Zicheng Liu, Ce~Liu, Michael Zeng, and Lijuan Wang.
\newblock Mm-react: Prompting chatgpt for multimodal reasoning and action.
\newblock {\em arXiv preprint arXiv:2303.11381}, 2023.

\bibitem{bai2025qwen3vltechnicalreport}
Shuai Bai, Yuxuan Cai, Ruizhe Chen, Keqin Chen, Xionghui Chen, Zesen Cheng, Lianghao Deng, Wei Ding, Chang Gao, Chunjiang Ge, Wenbin Ge, Zhifang Guo, Qidong Huang, Jie Huang, Fei Huang, Binyuan Hui, Shutong Jiang, Zhaohai Li, Mingsheng Li, Mei Li, Kaixin Li, Zicheng Lin, Junyang Lin, Xuejing Liu, Jiawei Liu, Chenglong Liu, Yang Liu, Dayiheng Liu, Shixuan Liu, Dunjie Lu, Ruilin Luo, Chenxu Lv, Rui Men, Lingchen Meng, Xuancheng Ren, Xingzhang Ren, Sibo Song, Yuchong Sun, Jun Tang, Jianhong Tu, Jianqiang Wan, Peng Wang, Pengfei Wang, Qiuyue Wang, Yuxuan Wang, Tianbao Xie, Yiheng Xu, Haiyang Xu, Jin Xu, Zhibo Yang, Mingkun Yang, Jianxin Yang, An~Yang, Bowen Yu, Fei Zhang, Hang Zhang, Xi~Zhang, Bo~Zheng, Humen Zhong, Jingren Zhou, Fan Zhou, Jing Zhou, Yuanzhi Zhu, and Ke~Zhu.
\newblock Qwen3-vl technical report, 2025.

\bibitem{zhu2025internvl3exploringadvancedtraining}
Jinguo Zhu, Weiyun Wang, Zhe Chen, Zhaoyang Liu, Shenglong Ye, Lixin Gu, Hao Tian, Yuchen Duan, Weijie Su, Jie Shao, Zhangwei Gao, Erfei Cui, Xuehui Wang, Yue Cao, Yangzhou Liu, Xingguang Wei, Hongjie Zhang, Haomin Wang, Weiye Xu, Hao Li, Jiahao Wang, Nianchen Deng, Songze Li, Yinan He, Tan Jiang, Jiapeng Luo, Yi~Wang, Conghui He, Botian Shi, Xingcheng Zhang, Wenqi Shao, Junjun He, Yingtong Xiong, Wenwen Qu, Peng Sun, Penglong Jiao, Han Lv, Lijun Wu, Kaipeng Zhang, Huipeng Deng, Jiaye Ge, Kai Chen, Limin Wang, Min Dou, Lewei Lu, Xizhou Zhu, Tong Lu, Dahua Lin, Yu~Qiao, Jifeng Dai, and Wenhai Wang.
\newblock Internvl3: Exploring advanced training and test-time recipes for open-source multimodal models, 2025.

\bibitem{kwon2023efficient}
Woosuk Kwon, Zhuohan Li, Siyuan Zhuang, Ying Sheng, Lianmin Zheng, Cody~Hao Yu, Joseph Gonzalez, Hao Zhang, and Ion Stoica.
\newblock Efficient memory management for large language model serving with pagedattention.
\newblock In {\em Proceedings of the 29th symposium on operating systems principles}, pages 611--626, 2023.

\bibitem{zheng2024sglang}
Lianmin Zheng, Liangsheng Yin, Zhiqiang Xie, Chuyue~Livia Sun, Jeff Huang, Cody~Hao Yu, Shiyi Cao, Christos Kozyrakis, Ion Stoica, Joseph~E Gonzalez, et~al.
\newblock Sglang: Efficient execution of structured language model programs.
\newblock {\em Advances in neural information processing systems}, 37:62557--62583, 2024.

\bibitem{aminabadi2022deepspeed}
Reza~Yazdani Aminabadi, Samyam Rajbhandari, Ammar~Ahmad Awan, Cheng Li, Du~Li, Elton Zheng, Olatunji Ruwase, Shaden Smith, Minjia Zhang, Jeff Rasley, et~al.
\newblock Deepspeed-inference: enabling efficient inference of transformer models at unprecedented scale.
\newblock In {\em SC22: International Conference for High Performance Computing, Networking, Storage and Analysis}, pages 1--15. IEEE, 2022.

\bibitem{patel2024splitwise}
Pratyush Patel, Esha Choukse, Chaojie Zhang, Aashaka Shah, {\'I}{\~n}igo Goiri, Saeed Maleki, and Ricardo Bianchini.
\newblock Splitwise: Efficient generative llm inference using phase splitting.
\newblock In {\em 2024 ACM/IEEE 51st Annual International Symposium on Computer Architecture (ISCA)}, pages 118--132. IEEE, 2024.

\bibitem{zhong2024distserve}
Yinmin Zhong, Shengyu Liu, Junda Chen, Jianbo Hu, Yibo Zhu, Xuanzhe Liu, Xin Jin, and Hao Zhang.
\newblock $\{$DistServe$\}$: Disaggregating prefill and decoding for goodput-optimized large language model serving.
\newblock In {\em 18th USENIX Symposium on Operating Systems Design and Implementation (OSDI 24)}, pages 193--210, 2024.

\bibitem{hu2024memserve}
Cunchen Hu, Heyang Huang, Junhao Hu, Jiang Xu, Xusheng Chen, Tao Xie, Chenxi Wang, Sa~Wang, Yungang Bao, Ninghui Sun, et~al.
\newblock Memserve: Context caching for disaggregated llm serving with elastic memory pool.
\newblock {\em arXiv preprint arXiv:2406.17565}, 2024.

\bibitem{qin2025mooncake}
Ruoyu Qin, Zheming Li, Weiran He, Jialei Cui, Feng Ren, Mingxing Zhang, Yongwei Wu, Weimin Zheng, and Xinran Xu.
\newblock Mooncake: Trading more storage for less computation—a $\{$KVCache-centric$\}$ architecture for serving $\{$LLM$\}$ chatbot.
\newblock In {\em 23rd USENIX Conference on File and Storage Technologies (FAST 25)}, pages 155--170, 2025.

\bibitem{dong2025hydrainfer}
Xianzhe Dong, Tongxuan Liu, Yuting Zeng, Liangyu Liu, Yang Liu, Siyu Wu, Yu~Wu, Hailong Yang, Ke~Zhang, and Jing Li.
\newblock Hydrainfer: Hybrid disaggregated scheduling for multimodal large language model serving.
\newblock {\em arXiv preprint arXiv:2505.12658}, 2025.

\bibitem{zhangspaceserve}
Shuoming Zhang, Jiacheng Zhao, Siqi Li, Xiyu Shi, Yangyu Zhang, Shuaijiang Li, Donglin Yu, Zheming Yang, Yuan Wen, Huimin Cui, et~al.
\newblock Spaceserve: Spatial multiplexing of complementary encoders and decoders for multimodal llms.
\newblock In {\em The Thirty-ninth Annual Conference on Neural Information Processing Systems}, 2025.

\bibitem{qiu2025modserve}
Haoran Qiu, Anish Biswas, Zihan Zhao, Jayashree Mohan, Alind Khare, Esha Choukse, {\'I}{\~n}igo Goiri, Zeyu Zhang, Haiying Shen, Chetan Bansal, et~al.
\newblock Modserve: Modality-and stage-aware resource disaggregation for scalable multimodal model serving.
\newblock {\em arXiv preprint arXiv:2502.00937}, 2025.

\bibitem{jia2025visualwebinstruct}
Yiming Jia, Jiachen Li, Xiang Yue, Bo~Li, Ping Nie, Kai Zou, and Wenhu Chen.
\newblock Visualwebinstruct: Scaling up multimodal instruction data through web search.
\newblock {\em arXiv preprint arXiv:2503.10582}, 2025.

\bibitem{chen2025sharegpt}
Junying Chen, Zhenyang Cai, Pengcheng Chen, Shunian Chen, Ke~Ji, Xidong Wang, Yunjin Yang, and Benyou Wang.
\newblock Sharegpt-4o-image: Aligning multimodal models with gpt-4o-level image generation.
\newblock {\em arXiv preprint arXiv:2506.18095}, 2025.

\end{thebibliography}

\end{CJK}
\end{document}